\documentclass[pdflatex,sn-basic]{sn-jnl}

\usepackage{graphicx}%
\usepackage{multirow}%
\usepackage{amsmath,amssymb,amsfonts}%
\usepackage{amsthm}%
\usepackage{mathrsfs}%
\usepackage[title]{appendix}%
\usepackage{xcolor}%
\usepackage{textcomp}%
\usepackage{manyfoot}%
\usepackage{booktabs}%
\usepackage{algorithm}%
\usepackage{algorithmicx}%
\usepackage{algpseudocode}%
\usepackage{listings}%

\theoremstyle{thmstyleone}%
\newtheorem{theorem}{Theorem}
\newtheorem{proposition}[theorem]{Proposition}%

\theoremstyle{thmstyleone}%
\newtheorem{lemma}{Lemma}%
\newtheorem{corollary}{Corollary}%
\newtheorem{hypothesis}{Hypothesis}%

\theoremstyle{thmstyletwo}%
\newtheorem{example}{Example}%
\newtheorem{remark}{Remark}%

\theoremstyle{thmstylethree}%
\newtheorem{definition}{Definition}%

\begin{document}

\title[Article Title]{Coarse-Grained Games: A Framework for Bounded Perception in Game Theory}


\author*[1]{\fnm{Takashi} \sur{Izumo}}\email{izumo.takashi@nihon-u.ac.jp}

\affil*[1]{\orgdiv{College of Law}, \orgname{Nihon University}, \orgaddress{\street{Kandamisakimachi}, \city{Chiyodaku}, \postcode{1018351}, \state{Tokyo}, \country{Japan}, ORCID: 0000-0003-0008-4729}}


\abstract{
In everyday life, we frequently make coarse-grained judgments. When we say that Olivia and Noah excel in mathematics, we disregard the specific differences in their mathematical abilities. Similarly, when we claim that a particular automobile manufacturer produces high-quality cars, we overlook the minor variations among individual vehicles. These coarse-grained assessments are distinct from erroneous or deceptive judgments, such as those resulting from student cheating or false advertising by corporations. 
Despite the prevalence of such judgments, little attention has been given to their underlying mathematical structure. In this paper, we introduce the concept of coarse-graining into game theory, analyzing games where players may perceive different payoffs as identical while preserving the underlying order structure. We call it a Coarse-Grained Game (CGG). This framework allows us to examine the rational inference processes that arise when players equate distinct micro-level payoffs at a macro level, and to explore how Nash equilibria are preserved or altered as a result.
Our key findings suggest that CGGs possess several desirable properties that make them suitable for modeling phenomena in the social sciences. This paper demonstrates two such applications: first, in cases of overly minor product updates, consumers may encounter an equilibrium selection problem, resulting in market behavior that is not driven by objective quality differences; second, the lemon market can be analyzed not only through objective information asymmetry but also through asymmetries in perceptual resolution or recognition ability.
}




\maketitle

\section{Introduction}\label{sec1}

\subsection{Background}\label{sec1.1}

Game theory has seen significant developments since its formalization by \cite{neumannmorgenstern1944}. The Folk Theorem, one of the fundamental results in repeated game theory, states that in infinitely repeated games, any feasible and individually rational payoff can be sustained as a Nash equilibrium \citep{friedman1971, abreu1988}, provided that players are sufficiently patient and that effective punishment strategies are available to deter deviations. This theorem assumes that players can accurately perceive differences in payoffs and respond accordingly.

However, in real-world scenarios, decision-makers often operate under rough perceptions, where information is aggregated, categorized, or subject to cognitive limitations. Consider a visit to a supermarket, where you find the section selling oranges. The display contains numerous oranges, all priced uniformly, but upon closer inspection, their sizes and sheen vary considerably. A person knowledgeable about oranges might think, ``This orange looks better than that one, so I'll choose it since they are the same price''. In contrast, someone less familiar with oranges may perceive them simply as ``two similar oranges'' and make no distinction between them. Now, suppose these two individuals must engage in a cooperative task that involves evaluating or selecting oranges. Their differing levels of perception could lead to misaligned expectations. Moreover, from the seller's perspective, an important strategic question arises: To what extent can irregular oranges be sold at the same uniform price? While if most customers are unable to distinguish orange quality, mixing in a larger proportion of lower-quality oranges may not pose a problem, if customers can accurately assess quality, a significant number of unsold, lower-quality oranges may remain. This example illustrates that while categorization is possible to some extent, the ability—or inability—to discern finer differences plays a crucial role in strategic interactions. It highlights how variations in perception can impact decision-making in economic transactions and cooperative settings.

On the contrary, such rough judgments are not only necessary but can also be efficient. It is important to recognize that they are inherently neutral—neither inherently good nor bad. This is particularly evident from the seller's perspective. For instance, a supermarket employee may find it more practical to display oranges of varying quality together at a uniform price rather than incur the cost of individually pricing each one. Instead of fearing that lower-quality oranges will be selectively avoided by buyers and remain unsold, it may be more efficient to accept this risk and simplify the pricing strategy. Thus, the fact that micro-level judgments are more precise does not imply that macro-level rough judgments should be avoided.

This necessitates an examination of the relationship between fine-grained judgments at the micro level and rough, aggregated judgments at the macro level. Such an inquiry is well-established in the natural sciences, where it is known as coarse-graining, a widely used methodology particularly in statistical mechanics and molecular chemistry \citep{Mori1965, Baschnagel2000, Reith2003, Marrink2004, Marrink2007, souza2021}. Coarse-graining is a modeling technique in which detailed, microscopic information is systematically simplified or aggregated into broader, macroscopic representations. This approach is employed when dealing with complex systems where tracking every individual component is computationally infeasible or practically unnecessary.

For example:
\begin{itemize}
\item In statistical mechanics, rather than describing the exact state of every individual particle, coarse-graining involves averaging over ensembles of particles to derive macroscopic properties such as temperature, pressure, and entropy.
\item In molecular chemistry, coarse-grained models replace groups of atoms with single interaction sites, significantly reducing computational complexity while preserving essential physical behaviors.
\end{itemize}

By applying a similar conceptual framework, we can analyze how coarse-graining in human decision-making allows individuals and institutions to manage complexity by filtering out fine details that are not operationally significant at a broader scale. To facilitate such an analysis, we seek to elucidate the mathematical structure of coarse-grained games (CGGs), providing a formal framework for understanding how strategic interactions evolve when players perceive payoffs at different levels of granularity.

\subsection{Comparison to Previous Research}\label{sec1.2}
Numerous studies have explored scenarios in which players do not possess perfect information or do not fully optimize \citep{simon1957, greenporter1984}. In repeated games, partial monitoring models relax the assumption of perfect observation, allowing each player to receive only noisy or aggregated signals about opponents' actions \citep{fudenberglevinemaskin1994, gossnertomala2006}. Meanwhile, bounded rationality approaches often restrict a player's cognitive capabilities or assumption-forming processes \citep{rubinstein1997, esponda2008}. Related lines of work also examine situations in which agents perceive or estimate payoffs only approximately, reflecting real-world constraints on decision-making \citep{lipton2003}.

Unlike many of these frameworks---which typically rely on incomplete information structures or probabilistic modeling---our study focuses on a coarse-grained perspective, wherein the granularity of each player's payoff partition can be explicitly controlled and need not be tied purely to uncertainty or noisy signals. By introducing a formalism for coarse partitions of payoff spaces, we offer an alternative lens for capturing cognitive resolution limitations. This coarse-graining view is thus closer to an ``information compression'' framework rather than a standard incomplete-information model, positioning our work as a complementary and more structural approach to understanding how limited discernibility can shape strategic outcomes.

The following list explains how CGGs differ from existing game-theoretic frameworks:

\begin{itemize}
\item CGGs are not a subclass of imperfect information games. In a CGG, players observe their opponent's strategies and payoffs through their own level of coarse perception. Since players are not assumed to be aware of differences in each other's levels of coarse perception, they subjectively perceive the game as one of perfect information. This differs from imperfect information games, where it is explicitly assumed that certain strategies or payoffs remain unobservable to the players \citep{osborne1994course}.
\item CGGs are not Bayesian games. In a Bayesian game \citep{harsanyi1967incomplete1, harsanyi1968incomplete2, harsanyi1968incomplete3}, each player possesses a probability distribution over types assigned by nature. In contrast, players in a CGG do not operate with such probabilistic beliefs but instead perceive the game through a coarse-grained payoff matrix (Section \ref{sec2.2}). In this structure, payoffs are not represented as precise real numbers but as sets of real values, making them fundamentally different from probability distributions. Note: Depending on the choice of strategy preprocessing methods as discussed in this paper (Section \ref{sec2.3}), a CGG could potentially be reformulated as a Bayesian game. However, this paper does not explore that possibility.
\item CGG is not a deception game \citep{sobe2020lying}. The fact that one player perceives information with a different level of granularity is not because they have been misled by someone else, nor is it the result of an opponent deliberately providing false information. For example, when a teacher tells a student, ``Your report was graded as excellent," instead of saying, ``Your report received a score of 93,'' they are neither lying nor spreading misinformation. The distinction lies in the coarseness of the information provided, not in its accuracy or truthfulness.
\item CGG does not concern naive players. In \cite{esponda2008} framework, naive players are those who fail to account for selection, leading to incorrect decision-making. However, in a CGG, the fundamental distinction lies between high-resolution players and low-resolution players, as will be discussed later (Section \ref{sec2.4}). While low-resolution players perceive payoffs with less clarity, this does not imply that they are naive. On the contrary, a low-resolution player can sometimes be the smarter choice in a practical sense. For example, instead of a teacher who meticulously distinguishes between a student who scored 91 and another who scored 92, adjusting their grading method to reflect this one-point difference, a teacher who simply categorizes both as ``excellent'' may be seen as more efficient and judicious.
\item CGG is related to information asymmetry but does not necessarily presuppose it. As shown in Section \ref{sec7.1}, even if the player who initially possesses more information voluntarily and actively shares it to eliminate the asymmetry, similar phenomena may still arise if the other party lacks the sensitivity to properly interpret that information. In this case, the asymmetry lies not in the amount of information available, but in the precision with which it is perceived. Therefore, CGGs cannot simply be reduced to a framework of information asymmetry; rather, they exhibit the characteristics of an epistemic framework.
\end{itemize}

\subsection{Structure of This Paper}\label{sec1.3}
The structure of this paper is as follows: Section \ref{sec2} defines CGGs from the perspective of coarse set theory, establishing a formal foundation for their analysis. Section \ref{sec3} interprets the meaning of payoffs in CGGs, introducing the distinction between subjective and objective payoffs. Section \ref{sec4} examines how Nash equilibria transform in CGGs, focusing on one-shot (non-repeated) strategic interactions. Section \ref{sec5} explores how changes in Nash equilibria can lead to unexpected incidental gains or losses for players in CGGs, highlighting the potential for unforeseen strategic outcomes. {Section \ref{sec6} analyzes infinitely repeated games, demonstrating that variations in perceived discount factors may prevent cooperation at discount rates that players would otherwise consider sufficient in coarse-grained payoff matrices. Section \ref{sec7} explores the applicability of CGGs in the social sciences by analyzing how consumers' coarse-grained perceptions influence decision-making. To illustrate this, we examine two cases: minor model changes in product markets and the lemon market. Section \ref{sec8} concludes the paper with a summary of key findings and directions for future research.

\section{Coarse-Grained Game}\label{sec2}
\subsection{Coarse Set}\label{sec2.1}
Based on \cite{izumo2025}, we construct a coarse-grained partition of totally ordered sets. First, we define the ordering among sets:

\begin{definition}[Element-wise Ordering]\label{def:element}
Suppose a set $S$ with a totally ordering $\preceq$. $\mathfrak{F}$ is a family of sets of $S$, in other words, $\mathfrak{F} \subseteq \mathcal{P}(S)$. Given two elements $A,B$ of $\mathfrak{F}$, if $x \in A$ and $y \in B$ can be compared regarding $(S, \preceq)$, then we denote it as $x \preceq_S y$ or $y \preceq_S x$. We define the element-wise ordering symbol $\preccurlyeq$ as follows:

\begin{equation}
\forall A,B \in \mathfrak{F}~ (\forall x \in A, \forall y \in B, x \preceq_S y \iff A \preccurlyeq B)
\end{equation}

\end{definition}

Now, we introduce the following axiom system.

\begin{enumerate}
\item $\forall x \in A~(x \preceq x)$
\item $\forall x,y \in A~ (x \preceq y \land y \preceq x \implies x = y)$
\item $\forall x,y,z \in A~ (x \preceq y \land y \preceq z \implies x \preceq z)$
\item $\forall x,y \in A~ (x \preceq y \lor y \preceq x)$
\item $\forall A \in \mathfrak{F}~ (A \preccurlyeq A)$
\item $\forall A,B \in \mathfrak{F}~ (A \preccurlyeq B \land B \preccurlyeq A \implies A = B)$
\item $\forall A,B,C \in \mathfrak{F}~ (A \preccurlyeq B \land B \preccurlyeq C \implies A \preccurlyeq C)$
\item $\forall A,B \in \mathfrak{F}~ (A \preccurlyeq B \lor B \preccurlyeq A)$
\end{enumerate}

\begin{remark}
For a detailed discussion on why expressions like \(\max, \min, \sup,\) or \(\inf\) may fail in coarse-grained partitions—particularly over the real numbers—see \cite{izumo2025}. In brief, open intervals such as \((a,b)\) do not admit true maximum or minimum elements, so rewriting  
\[
  \forall x \in A, \forall y \in B,\ x \le y
\]  
as \(\max(A) \le \min(B)\) (or \(\sup(A) \le \inf(B)\)) can break down whenever endpoints fall outside the sets in question.
\end{remark}

Under this axiom system, we define a coarse-grained partition.

\begin{definition}[Coarse-Grained Partition]\label{partition}
Suppose a totally ordered set $(U,\preceq)$. $U$ serves as the underlying set for a family of sets $\mathfrak{F} \subseteq \mathcal{P}(U)$. If the element-wise ordering is applicable to $\mathfrak{F}$ to make $(\mathfrak{F},\preccurlyeq)$ and $\mathfrak{F}$ satisfies the following condition, we call it a \textit{coarse-grained partition} of $(U,\preceq)$ and denote it as $\mathfrak{G}$ of which elements are reffered to \textit{grains} of $(U,\prec)$.

\begin{equation}
\forall u \in U, \exists! G \in \mathfrak{F}~ (u \in G)
\end{equation}
\end{definition}

\begin{remark}
\cite{izumo2025} states that whether \( G \) is merely a set, a set in which all elements are equivalent, or a totally ordered set can be determined based on the research objective. In this paper, since \( G \) must be treated as both an open and a closed set of real numbers, we assume that \( G \) is a totally ordered set.
\end{remark}

Additionally, we define the collection of the coarse-grained partitions of a totally ordered set.

\begin{definition}[Collection of Coarse-Grained Partitions]\label{collection} Suppose a set that contains all coarse-grained partitions of a totally ordered set $(U,\preceq)$. We denote it as $\mathcal{C}(U,\preceq)$.
\end{definition}

\begin{example}
Given a set $\{1,2,3\}$ with the natural order $\leq$, 

\begin{align*}
\mathcal{C}(U, & \leq) = \{(\mathfrak{G}_1, \preccurlyeq_1), (\mathfrak{G}_2, \preccurlyeq_2), (\mathfrak{G}_3, \preccurlyeq_3), (\mathfrak{G}_4, \preccurlyeq_4)\} \\
= & \{ \\
& (\mathfrak{G}_1, \preccurlyeq_1) = \\
& \quad (\{\{\{1\},\{2\},\{3\}\}, \{(\{1\},\{1\}),(\{1\},\{2\}),(\{1\},\{3\}),(\{2\},\{2\}),(\{2\},\{3\}),(\{3\},\{3\})\}), \\
& (\mathfrak{G}_2, \preccurlyeq_2) = \\
& \quad (\{\{1\},\{2,3\}\}, \{(\{1\},\{1\}),(\{1\},\{2,3\}),(\{2,3\},\{2,3\})\}), \\
& (\mathfrak{G}_3, \preccurlyeq_3) = \\
& \quad (\{\{1,2\},\{3\}\}, \{(\{1,2\},\{1,2\}),(\{1,2\},\{3\}),(\{3\},\{3\})\}), \\
&  (\mathfrak{G}_4, \preccurlyeq_4) = \\
& \quad (\{\{1,2,3\}, \{(\{1,2,3\},\{1,2,3\})\}) \\ 
& \}.
\end{align*} 
\end{example}

\begin{remark}
Since this paper deals with payoff matrices in game theory, the underlying set \( U \) is taken to be the set of real numbers \( \mathbb{R} \). The total order \( \preceq \) on \( U \) is naturally given by the standard order relation on real numbers, i.e., \( \leq \). 
\end{remark}

\subsection{Coarse-Grained Payoff Matrix}\label{sec2.2}
Applying coarse set theory, we define the concept of a coarse-grained payoff matrix.

\begin{definition}[Coarse-Grained Payoff Matrix]\label{cg-payoff-matrix}
Let \( (\mathbb{R}, \leq) \) be a totally ordered set of payoffs, and let \( \mathfrak{G} \) be a coarse-grained partition of \( (\mathbb{R}, \leq) \), denoted by  
\[
\mathfrak{G} = \{G_\iota\}_{\iota \in I}, \quad \text{where } \bigcup_{\iota \in I} G_\iota = \mathbb{R} \text{ and } G_{\iota} \cap G_{\iota'} = \emptyset \text{ for } \iota \neq \iota'.
\]
A Coarse-Grained Payoff Matrix (CGPM) under partition \( \mathfrak{G} \) is a payoff matrix \( M \) in which each cell \( M_{i_1,i_2,\dots,i_n} (n \in \mathbb{N}) \) contains payoffs mapped to their respective grains in \( \mathfrak{G} \). Formally,  
\[
M_{i_1,i_2,\dots,i_n} \in \mathfrak{G}, \quad \forall i.
\]
\end{definition}

\begin{example}\label{coarse-grained-matrix}
Given a coarse-grained partition $\mathfrak{G}$, 
\begin{equation*}
\mathfrak{G} = \{\dots, [-6,-4), [-4,-2), [-2,0), \{0\}, (0,2], (2,4], (4,6], \dots\}.
 \end{equation*}

\noindent 
An example of the coarse-grained payoff matrix is:

\[
\begin{array}{c|cc}
k_1 \backslash k_2& \textbf{Cooperate} & \textbf{Defect} \\ \hline
\textbf{Cooperate} & ((0,2], \, (2,4]) & (\{0\},\, [-6,-4)) \\
\textbf{Defect}    & ([-8,-6),\, \{0\}) & ([-4,-2), \, [-6,-4))
\end{array}
\]
\end{example}

\subsection{Entropy-Maximizing Preprocessing}\label{sec2.3}
When the payoff matrix contains sets of numerical values instead of single numerical payoffs, players cannot directly compute their optimal actions. Players must first determine how to handle these multi-valued payoffs to make decisions. This preliminary decision-making process is referred to as a strategy preprocessing in this paper, and multiple approaches can be considered.

One possible strategy preprocessing is to treat options with multi-valued payoffs as indeterminate and exclude them from consideration (ignore strategy preprocessing). Under this approach, in the given example, players must give up, in other words, the game itself would become ill-defined and unplayable.

However, rather than resorting to such a thought-stopping approach, a more rational strategy preprocessing should involve a systematic method for uniquely transforming multi-valued payoffs into single numerical values. There are multiple ways to achieve this, but one theoretically and empirically justifiable method is to apply the principle of maximum entropy, ensuring that the transformation is carried out to preserve as much information as possible while enabling strategic decision-making.

\begin{definition}[Entropy-Maximizing Preprocessing]\label{entropy}
If a payoff matrix contains some sets of numerical values, assume that each value within the set occurs with equal probability. 

\begin{itemize}
\item If a set contains only a single real number as its element, it is replaced by that real number.
\item If the set is a finite discrete set, compute the arithmetic mean (expected value under a uniform distribution) and replace the set with this computed expectation.  
\item If the set is a continuous interval \( (a,b) \), $(a,b]$, $[a,b)$, or \( [a,b] \), assume a uniform probability density function over the interval and compute the expected value:
\end{itemize}
\begin{equation*}
E[(a,b)] = \frac{1}{b-a} \int_a^b x \,dx = \frac{a+b}{2}.
\end{equation*}
This approach is referred to as the Entropy-Maximizing Preprocessing (EMP). Note that the open set \((a,b)\) should not be confused with the tuple of two payoffs \((a,b)\).
\end{definition}

\begin{example}
Applying the EMP to Example \ref{coarse-grained-matrix}:
\[
\begin{array}{c|cc}
k_1 \backslash k_2 & \textbf{Cooperate} & \textbf{Defect} \\ \hline
\textbf{Cooperate} & (1,\, 3) & (0,\, -5) \\ 
\textbf{Defect}    & (-7,\, 0) & (-3,\, -5)
\end{array}
\]
\end{example}

The principle of maximum entropy can theoretically justify the EMP. According to this principle, when assigning probabilities in the absence of additional information, the most unbiased and least assumption-laden distribution is the one that maximizes entropy \citep{jaynes1957}. In this context, this means that if a payoff is given as a set of numerical values, the most objective way to interpret it is to assume a uniform probability distribution over its elements. By doing so, we ensure that no particular value is favored without justification, and we obtain a single representative payoff by computing the expected value. This transformation allows players to make rational decisions while preserving as much of the original information as possible.

Furthermore, the EMP can also be justified empirically. Consider a consumer who evaluates apples based on four categories: ``sweet'', ``moderately sweet'', ``slightly sweet'', and ``not sweet''. Within this framework, subtle differences in sugar content among apples categorized as ``sweet'' become indistinguishable. Even if a precise measurement using a refractometer could reveal the exact sugar content distribution within this category—showing which levels of sweetness appear more frequently---the consumer, lacking such detailed information, has no choice but to assume a uniform probability distribution across all apples labeled as ``sweet''. In other words, despite the existence of an underlying variation, the absence of finer perception forces the consumer to treat all apples in this category as occurring with equal likelihood.

\subsection{Cognitive Resolution}\label{sec2.4}
\subsubsection{Basic Idea}

From a cognitive perspective, a CGPM reflects an individual's ability to understand and evaluate reality. In this paper, we refer to this capability as resolution. The analogy of cameras inspires this term: just as a high-resolution camera captures fine details that a low-resolution camera cannot, a high-resolution player perceives a payoff matrix with finer grains, whereas a low-resolution player perceives a payoff matrix with coarser grains. 

\begin{remark}
The term ``resolution'' is typically a technical term and may not be commonly used in the context of human cognition. However, the cognitive state under discussion in this paper is neither misunderstanding nor misperception. The phrase misunderstanding is ambiguous in that it does not clearly specify what the players disagree on, and it may also imply that one player is naive. Meanwhile, misperception can encompass cases where one mistakes something nonexistent for real, making it too broad for our intended meaning.  In light of this, the author draws attention to the fact that in Japanese, the term \textit{kaizōdo} (in English: resolution) is commonly used in a non-technical sense to describe situations where a person fails to recognize subtle distinctions in an object or concept; for example, when a friend says, ``That actor and this actor are both equally excellent at acting'', it is natural in Japanese to respond, ``Your resolution is too low—there is actually a difference in their acting skills''.  Given its appropriateness in conveying the intended notion of coarse-grained cognition, this paper adopts resolution as the preferred term.
\end{remark}

To compare and analyze resolution levels, we define the $\mathfrak{G}$-resolution payoff matrix \( M^{\mathfrak{G}} \) as follows.

\begin{definition}[$\mathfrak{G}$-Resolution Payoff Matrix]\label{g-res-matrix}
Suppose the following payoff matrix:
\begin{align*}
& M^{\text{base}}_{j_1,j_2,\dots,j_k\dots,j_n} = (x_1, x_2, \dots,x_k,\dots,x_n), \\ 
& \quad \text{where } j,k,n \in \mathbb{N}, x_k \in \mathbb{R} \text{ for all } k, \text{ and } j_k \text{ corresponds a strategy of player } k.
\end{align*}

\noindent
This is a standard payoff matrix—in other words, the format that first comes to mind when we think of a payoff matrix.

Given a collection of coarse-grained partitions \( \mathcal{C}(\mathbb{R},\leq) = \{(\mathfrak{G}_\theta, \preccurlyeq_\theta) \mid \theta \in \Theta \} \), the $\mathfrak{G}$-resolution payoff matrix \( M^{\mathfrak{G}} \) is defined as follows:
\[
M^{\text{base}}_{j_1,j_2,\dots, j_k,\dots,j_n} = (G_1, G_2, \dots, G_k, \dots, G_n), \quad \text{where } G_k \in \mathfrak{G} \text{ for all } k.
\]
\end{definition}

\begin{proposition}[Uniqueness of the Finest-Resolution Payoff Matrix]\label{finest-res-matrix}
Given a collection of coarse-grained partitions \( \mathcal{C}(\mathbb{R},\leq) = \{(\mathfrak{G}_\theta,\preccurlyeq_\theta) \mid {\theta \in \Theta} \} \), the finest-resolution payoff matrix \( M^{\mathfrak{G}^\text{finest}} \) is defined as follows:
\begin{align*}
& M^{\mathfrak{G}^\text{finest}}_{j_1,j_2,\dots,j_k,\dots,j_n} = (\{x_1\}, \{x_2\}, \dots,\{x_k\},\dots,\{x_n\}), \\
& \quad \text{where } j,k,n \in \mathbb{N} \text{ and } x_k \in \mathbb{R} \text{ for all } k.
\end{align*}

\noindent
The coarse-grained partition of $(\mathbb{R},\leq)$ associated with \( M^{\mathfrak{G}^\text{finest}} \) is:
\[
\mathfrak{G}^{\text{finest}} = \{\{x\} \mid x \in \mathbb{R} \}.
\]

The finest-resolution payoff matrix $M^{\mathfrak{G}^\text{finest}}$ is uniquely determined.
\end{proposition}

\begin{proof}[Proof of Proposition \ref{finest-res-matrix}]
By definition, the finest-grained partition $\mathfrak{G}^{\text{finest}}$ is given by:
   \[
   \mathfrak{G}^{\text{finest}} = \{ \{x\} \mid x \in \mathbb{R} \}.
   \]
This means that each element of $U$ is treated as a singleton set. $\mathfrak{G}^{\text{finest}}$ satisfies the partition conditions, that is, covering property: It holds that
     \[
     \bigcup_{G \in \mathfrak{G}^{\text{finest}}} G = \mathbb{R},
     \]

\noindent
and disjointness: For any $u, v \in \mathbb{R}$ where $u \neq v$, we have
     \[
     \{u\} \cap \{v\} = \emptyset.
     \]

The finest-resolution payoff matrix $M^{\mathfrak{G}^\text{finest}}$ is defined based on $\mathfrak{G}^{\text{finest}}$, where each entry is of the form:
\begin{align*}
& M^{\mathfrak{G}^\text{finest}}_{j_1,j_2,\dots,j_k,\dots,j_n} = (\{x_1\}, \{x_2\}, \dots,\{x_k\},\dots,\{x_n\}), \\
& \quad \text{where } j,k,n \in \mathbb{N} \text{ and } x_k \in \mathbb{R} \text{ for all } k.
\end{align*}

Suppose there exists another finest-resolution payoff matrix $M^{\mathfrak{G}^{\text{finest}'}}$. Then, its corresponding partition $\mathfrak{G}^{\text{finest}'}$ must be different from $\mathfrak{G}^{\text{finest}}$. However, $\mathfrak{G}^{\text{finest}}$ is uniquely determined by $(\mathbb{R},\leq)$, so it is impossible for $\mathfrak{G}^{\text{finest}} \neq \mathfrak{G}^{\text{finest}'}$. This contradiction implies that $M^{\mathfrak{G}^{\text{finest}}}$ is uniquely determined.

Thus, the proof is complete.
\end{proof}

\begin{corollary}[Lowest-Resolution Payoff Matrix]\label{lowest-res-matrix}
The lowest-resolution payoff matrix $M^{\mathfrak{G}^\text{lowest}}$ is defined as follows:
\begin{align*}
& M^{\mathfrak{G}^\text{lowest}}_{j_1,j_2,\dots,j_k,\dots,j_n} = (A_1, A_2, \dots, A_k, \dots, A_n), \\
& \quad \text{where } j,k,n \in \mathbb{N} \text{ and } A_k = \mathbb{R} \text{ for all } k.
\end{align*}

\noindent
The coarse-grained partition of $(\mathbb{R},\leq)$ associated with \( M^{\mathfrak{G}^{\text{lowest}}} \) is:
\[
\mathfrak{G}^{\text{lowest}} = \{\mathbb{R} \}.
\]

The lowest-resolution payoff matrix $M^{\mathfrak{G}^{\text{lowest}}}$ is uniquely determined.
\end{corollary}

\subsection{Coarse-Grained Game}\label{sec2.5}
\subsubsection{Definition}

\begin{definition}[Coarse-Grained Game]\label{cg-game}
A Coarse-Grained Game \( \mathcal{G} \) is defined as a tuple:

\[
\mathcal{G} = (N, S, u, \mathfrak{G}, \Phi, \Psi)
\]

where:
\begin{itemize}
\item $N = \{1, 2, \dots, k, \dots, n\}~ (k,n \in \mathbb{N})$ is a finite set of players. Note: Among the following sets and tuples, some have the same number of elements as the player set \( N \) (i.e., those where the indixes \( k \) and $n$ are valid), while others do not.
\item \( S = (S_1, S_2, \dots, S_k, \dots, S_n) \) is the tuple of the sets of strategies for each player $k$, where $S_k = \{s_1,s_2,\dots,s_j, \dots,s_m\}~ (j,m \in \mathbb{N}).$
\item The tuple \( u = (u_1, u_2, \dots,u_k,\dots,u_n) \) represents the payoff functions for each player $k$ in a game. Each function \( u_k \) assigns a real number payoff based on the strategies chosen by all players. Formally, 
\begin{equation*}
u_k: S_1 \times S_2 \times \dots \times S_k \times \dots \times S_n \to \mathbb{R},
\end{equation*}
where \( S_k \) denotes the strategy set of player \( k \). For a specific strategy profile
\begin{equation*}
(s_{j_1}, s_{j_2}, \dots, s_{j_k}, \dots, s_{j_n}), \quad \forall s_{j_k} \in S_k = \{s_1, \dots, s_j, \dots, s_n \},
\end{equation*}
the payoff to player \( k \) is given by \( u_k (s_{j_1}, s_{j_2}, \dots, s_{j_n}) \).
\item \( \mathfrak{G} = (\mathfrak{G}_1, \mathfrak{G}_2, \dots, \mathfrak{G}_k, \dots, \mathfrak{G}_n) \) is the tuple of coarse-grained partitions, where each \( \mathfrak{G}_k \) is assigned to player \( k \). In this paper, $\mathfrak{G}_k$ is of $(\mathfrak{G}_k, \leq_k) \in \mathcal{C}(\mathbb{R}, \leq)$.
\item \( \Phi = (\varphi_1, \varphi_2, \dots, \varphi_k, \dots, \varphi_n) \) is the tuple of coarse-graining functions, where each \( \mathfrak{G}_k \) is assigned to player \( k \):
\begin{align*}
& \varphi_k: \mathbb{R} \to \mathfrak{G}_k; \\
& \quad\quad \varphi_k(x) = G, \quad \text{where } x \in G \text{ and } G \in \mathfrak{G}_k.
\end{align*}
    
This means that each player's payoff $u_k(s_{j_1}, s_{j_2}, \dots, s_{j_n})$ is mapped to the corresponding element in their assigned coarse-grained partition $\mathfrak{G}_k$.
\item \( \Psi = (\psi_1,\psi_2,\dots,\psi_k,\dots,\psi_n) \) is the tuple of preprocessing methods for each player, e.g., EMP. Strategy preprocessing functions represent additional transformations or modifications applied to the strategy space before the game is played, according to the following rule:
        \[
        \psi_k: \mathfrak{G}_k \to \mathbb{R}; \quad \psi_k(G) = y,
        \]
with $y$ determined based on the specific preprocessing employed by $k$. In this paper, we assume that all players employ the EMP, and we denote their strategy as $\psi_{\text{EMP}}$ (i.e., $\psi_k = \psi_{\text{EMP}}$ for all $k$).
\end{itemize}
\end{definition}

\begin{remark}
The function \( \Phi \) is a tuple of coarse-graining functions, where each \( \varphi_k \) applies a player-specific transformation to payoffs. If the function is to be applied to the entire matrix rather than to individual payoffs, it is acceptable to write \(\Phi_k(M)\), provided that no confusion arises. \( \Phi_k \) maps the detailed payoffs to a coarser representation based on the assigned partition \( \mathfrak{G}_k \):
\begin{align*}
& \Phi_k(M_{j_1, j_2, \dots, j_k, \dots, j_n}) = M'_{j_1, j_2, \dots, j_n}, \\
& \quad \text{ where } (j \in \mathbb{N}), \text{ and } j_k \text{ corresponds to a strategy of } k. 
\end{align*}
In this case, $\Phi_k$ functions:
\[
(\mathbb{R}^n)^{|S_1| \times |S_2| \times \dots \times |S_k| \times \dots \times |S_n|}
\to (\mathcal{P}(\mathbb{R})^n)^{|S_1| \times |S_2| \times \dots \times |S_k| \times \dots \times |S_n|}.
\]
This means that each player's original payoff is mapped to a coarser value according to their own partitioning scheme. Similarily:
\[
\Psi_k (M'_{j_1, j_2, \dots, j_n}) = M''_{j_1, j_2, \dots, j_n}~ (j \in \mathbb{N}).
\]
In this case, $\Psi_k$ functions:
\[
(\mathcal{P}(\mathbb{R})^n)^{|S_1| \times |S_2| \times \dots \times |S_n|}
\to (\mathbb{R}^n)^{|S_1| \times |S_2| \times \dots \times |S_n|}.
\]
\end{remark}

In the following, we refer to the payoff matrix constructed to apply \( \varphi_k \) as \( M^{\text{base}} \) which is determined by the combination of $N$, $S$, and $u$. The coarse-grained payoff matrix obtained by player \( k \) through this application is denoted as \(  M^{\mathfrak{G}_k}_k \). Since the subscription of a coarse-grained payoff matrix is equal to the subscription of $\mathfrak{G}$, we omit the former as $M^{\mathfrak{G}_k}$. Applying \( \psi_k \) to \( M^{\mathfrak{G}_k} \) yields the final payoff matrix of each player, which we denote as \( M'_k \).

\subsubsection{Example: Coarse-Grained Prisoners Dilemma}
\paragraph{The Standard Payoff Matrix \( M^{\text{base}} \) (Standard Prisoner's Dilemma)}

\[
\begin{array}{c|cc}
player 1 ~\backslash~ player 2 & \textbf{Remain Silent} & \textbf{Confess} \\ \hline
\textbf{Remain Silent} & (-1, -1) & (-5, 0) \\  
\textbf{Confess} & (0, -5) & (-3, -3)  
\end{array}
\]

\paragraph{Coarse-Grained Payoff Matrix for Player 1 (\( M^{\mathfrak{G}_1} \))}

Player 1 has the following coarse-grained partition:  
\[
\mathfrak{G}_1 = \{\dots, [-6,-4), [-4,-2), [-2,0), \{0\}, (0,2], (2,4], (4,6], \dots \}
\]
Applying this partition to the base payoff matrix $M^{\text{base}}$ through $\varphi_{1}$:

\[
\begin{array}{c|cc}
player 1 ~\backslash_{(\mathfrak{G}_1)}~ player 2 & \textbf{Remain Silent} & \textbf{Confess} \\ \hline
\textbf{Remain Silent} & ([-2,0), [-2,0)) & ([-6,-4), \{0\}) \\  
\textbf{Confess} & (\{0\}, [-6,-4)) & ([-4,-2), [-4,-2))  
\end{array}
\]

\paragraph{Coarse-Grained Payoff Matrix for Player 2 (\( M^{\mathfrak{G}_2} \))}

Player 2 has the following coarse-grained partition:  
\[
\mathfrak{G}_2 = \{\dots, [-18,-12), [-12,-6), [-6,0), \{0\}, (0,6], (6,12], (12,18],\dots \}
\]
Applying this partition to $M^{\text{base}}$ through $\varphi_2$:

\[
\begin{array}{c|cc}
player 1 ~\backslash_{(\mathfrak{G}_2)}~ player 2 & \textbf{Remain Silent} & \textbf{Confess} \\ \hline
\textbf{Remain Silent} & ([-6,0), [-6,0)) & ([-6,0), \{0\}) \\  
\textbf{Confess} & (\{0\}, [-6,0)) & ([-6,0), [-6,0))  
\end{array}
\]

\begin{remark}
The following table shows how each value in \( M^{\text{base}} \) is mapped to the coarse-grained payoff matrix \( M^{\mathfrak{G}_1} \) according to the partition \( \mathfrak{G}_1 \):

\[
\begin{array}{c|c}
M^{\text{base}} \text{ Value} & \mathfrak{G}_1 \text{ Mapping} \\ \hline
-1  & [-2,0) \\
-5  & [-6,-4) \\
0  & \{0\} \\
-3  & [-4,-2)
\end{array}
\]

The following table shows how each value in \( M^{\text{base}} \) is mapped to the coarse-grained payoff matrix \( M^{\mathfrak{G}_2} \) according to the partition \( \mathfrak{G}_2 \):

\[
\begin{array}{c|c}
M^{\text{base}} \text{ Value} & \mathfrak{G}_2 \text{ Mapping} \\ \hline
-1  & [-6,0) \\
-5  & [-6,0) \\
0  & \{0\} \\
-3  & [-6,0)
\end{array}
\]
\end{remark}

We apply the EMP to both the matrices through $\Psi$ and obtain $M'_1$ (above) and $M'_2$ (below):

    \[
    \begin{array}{c|cc}
    player 1 ~\backslash_{(\mathfrak{G}_1 + \text{EMP})}~ player 2 & \textbf{Remain Silent} & \textbf{Confess} \\ \hline
    \textbf{Remain Silent} & (-1, -1) & (-5, 0) \\  
    \textbf{Confess}    & (0, -5) & (-3, -3)  
    \end{array}
    \]

    \[
    \begin{array}{c|cc}
    player 1 ~\backslash_{(\mathfrak{G}_2 + \text{EMP})}~ player 2 & \textbf{Remain Silent} & \textbf{Confess} \\ \hline
    \textbf{Remain Silent} & (-3, -3) & (-3, 0) \\  
    \textbf{Confess}    & (0, -3) & (-3, -3)  
    \end{array}
    \]

\begin{remark}
In this paper, any additional operations applied to the payoff matrix, such as the perspective of coarsening or preprocessing, are denoted using a subscript to the right of the backslash symbol.
\end{remark}

\section{Interpreting Coarse-Grained Payoffs}\label{sec3}
\subsection{Objective vs. Subjective Payoffs in Coarse-Grained Games}\label{sec3.1}

Now that we have each player's coarse-grained payoff matrix \( M^{\mathfrak{G}_k} \) and its preprocessed version $M'_k$, we need to establish a clear framework for interpretation and computation.

\begin{definition}[Objective Payoff and Subjective Payoff]\label{payoff}
In the base matrix, the payoff that player \( k \) receives is referred to as player \( k \)'s objective payoff. Meanwhile, in the matrix obtained by applying preprocessing to player \( k \)'s coarse-grained matrix, the payoff that player \( l \) receives is called player \( l \)'s subjective payoff from the perspective of player \( k \).
\end{definition}

\begin{itemize}
    \item \textbf{Objective Payoff within \( M^{\text{base}} \)}:  
    The base payoff matrix represents the true underlying structure of the game. This matrix is independent of player perception and reflects the game's fundamental incentives. However, in a CGG, players do not have direct access to \( M^{\text{base}} \).

    \item \textbf{Subjective Payoff within \( M'_k \)}:  
    The payoff matrix perceived by player \( k \) after applying their coarse-grained partition \( \mathfrak{G}_k \) and their strategy preprocessing methods. Different players may have different payoff matrices, depending on their resolution level. Since players make decisions based on their perceived payoffs, \( M'_k \) is the basis for each player's strategy optimization.
\end{itemize}

Since players act based on their subjective perception of payoffs, the Nash equilibrium computation should use \( M_k \). However, once the Nash equilibrium strategies are found, we can compare them to what would happen in \( M^{\text{base}} \).

In this paper, to distinguish between the Nash equilibrium strategy pair in the base payoff matrix and the equilibrium strategy pair in each player's coarse-grained payoff matrix, we introduce the notation \( \dag \) for the latter. 

\begin{itemize}
\item The best response of player $k$ in the base matrix is denoted as \( s^*_k \).  
\item The best response of player $k$ in the coarse-grained matrix as perceived by player \( l \) is denoted as \( s^{\dag_l}_k \). From player \( k \)'s own perspective, their optimal strategy is \( s^{\dag_k}_k \).
\item For mixed strategies, the best response of player $k$ in the base matrix is denoted by the probability vector $p^*_k = (p^*_{1_k}, p^*_{2_k}, \dots, p^*_{j_k}, \dots, p^*_{m_k})$, where $p^*_{j_k}$ represents the probability distribution that player $k$ selects strategy $s_j$ from their strategy set $S_k = \{s_1,s_2,\dots,s_j,\dots,s_m\}$, with the constraints:
\[
\sum^m_{j=1} p^*_{j_k} = 1 \text{ and } 0 \leq p^*_{j_k} \leq 1 \quad \forall j,k,m \in \mathbb{N}.
\]
\item Similarly, the best response of player \( k \) in the coarse-grained matrix as perceived by player \( l \) is denoted by \( p^{\dag_l}_k = (p^{\dag_l}_{1_k}, p^{\dag_l}_{2_k}, \dots, p^{\dag_l}_{j_k}, \dots, p^{\dag_l}_{m_k}) \), where \( p^{\dag_l}_{j_k} \) represents the probability that player \( k \) selects strategy \( s_j \) from their strategy set \( S_k \), with the constraints:
\[
\sum^m_{j=1} p^{\dag_l}_{j_k} = 1 \text{ and } 0 \leq p^{\dag_l}_{j_k} \leq 1 \quad \forall j,k,l,m \in \mathbb{N},
\]
\end{itemize}

This notation clearly differentiates between the objective best responses in the base game and the subjective best responses in the coarse-grained perspectives of each player.

\subsection{Application to Coarse-Grained Prisoner's Dilemma}\label{sec3.2}
\subsubsection{Preparation}\label{sec3.2.1}

Let us analyze the Nash equilibrium of the coarse-grained Prisoner's Dilemma as a two-player one-shot game with pure strategies and the EMP from three different perspectives: the objective view (Table \ref{tb:base}), player 1's perspective (Table \ref{tb:m1}), and player 2's perspective (Table \ref{tb:m2}).  

\begin{itemize}
\item From the objective perspective, the conclusion is well known: the Nash equilibrium is (Confess, Confess), where each player receives a payoff of $-3$ (equivalent to a 3-year prison sentence).  
\item From player 1's perspective, despite their resolution being coarser than the objective view, they arrive at the same equilibrium outcome as in the standard game.  
\item From player 2's perspective, however, the game becomes less decisive, leading to equilibrium selection problem.
\end{itemize}

\begin{table}[h]
    \centering
    \caption{The Base Payoff Matrix}
    \label{tb:base}
    \begin{tabular}{c|cc}
        $player 1 ~\backslash~ player 2$ & \textbf{Remain Silent} & \textbf{Confess} \\ \hline
        \textbf{Remain Silent} & $(-1, -1)$ & $(-5, 0)$ \\  
        \textbf{Confess}    & $(0, -5)$ & $(-3, -3)$  
    \end{tabular}
\end{table}

\begin{table}[h]
    \centering
    \caption{EMP-Applied Payoff Matrix for Player 1 ($ M_1' $)}
    \label{tb:m1}
    \begin{tabular}{c|cc}
        $player1 ~\backslash_{(\mathfrak{G}_1+\text{EMP})}~ player2$ & \textbf{Remain Silent} & \textbf{Confess} \\ \hline
        \textbf{Remain Silent} & $([-2,0), [-2,0))$ $\to$ $(-1, -1)$ & $([-6,-4), \{0\})$ $\to$ $(-5, 0)$ \\  
        \textbf{Confess}    & $(\{0\}, [-6,-4))$ $\to$ $(0, -5)$ & $([-4,-2), [-4,-2))$ $\to$ $(-3, -3)$  
    \end{tabular}
\end{table}

\begin{table}[h]
    \centering
    \caption{EMP-Applied Payoff Matrix for Player 2 ($ M_2' $)}
    \label{tb:m2}
    \begin{tabular}{c|cc}
        $player1 ~\backslash_{(\mathfrak{G}_2+\text{EMP})}~ player2$ & \textbf{Remain Silent} & \textbf{Confess} \\ \hline
        \textbf{Remain Silent} & $([-6,0),[-6,0))$ $\to$ $(-3, -3)$ & $([-6,0),\{0\})$ $\to$ $(-3, 0)$ \\  
        \textbf{Confess}    & $(\{0\},[-6,0))$ $\to$ $(0, -3)$ & $([-6,0),[-6,0))$ $\to$ $(-3, -3)$  
    \end{tabular}
\end{table}

\begin{remark}
The players 1 and 2 are not assigned their resolution levels arbitrarily; rather, they reflect realistic psychological tendencies that could influence how individuals perceive sentencing:

\begin{itemize}
\item Player 1: This player perceives the difference in suffering every additional two years in prison brings, meaning they distinguish between significant sentencing gaps. However, when the sentence is only one or two years, they tend to view it as a single, undifferentiated category rather than separate punishments.  
\item Player 2: This player is an all-or-nothing thinker and perceives the situation in absolute terms. To them, the only meaningful distinction is between being guilty and not guilty—they do not distinguish between different sentence lengths.
\end{itemize}
\end{remark}

\subsubsection{Process of Reasoning}

Let us now analyze what is happening to each player from their subjective perspective.  

\paragraph{Player 1's Perspective:}

\begin{itemize}
\item Player 1 observes the game through their coarse-grained matrix with the EMP, that is, \( M'_1 \) and applies standard game-theoretic reasoning to derive their Nash equilibrium.  
\item Based on \( M'_1 \), the expected Nash equilibrium \( (s^{\dag_1}_{2_1}, s^{\dag_1}_{2_2}) \) is (Confess, Confess), yielding an expected payoff of $(-3, -3)$. 
\end{itemize}

\paragraph{Player 2's Perspective:}

\begin{itemize}
\item Player 2, on the other hand, makes decisions based on \( M'_2 \).  
\item Given \( M'_2 \), the Nash equilibrium appears to be in \( (s^{\dag_2}_{2_1}, s^{\dag_2}_{2_2}) \), \( (s^{\dag_2}_{1_1}, s^{\dag_2}_{2_2}) \), and \( (s^{\dag_2}_{2_1}, s^{\dag_2}_{1_2}) \), namely (Confess, Confess), (Remain Silent, Confess), and (Confess, Remain Silent). This suggests that while a strictly dominant strategy exists from the perspective of player 1, player 2 now faces an equilibrium selection problem due to the multiple Nash equilibria.
\end{itemize}

The concept of a focal point, introduced by \cite{schelling1960strategy}, refers to a solution or equilibrium that players are naturally inclined to choose due to its psychological salience, symmetry, or social convention, even in the absence of explicit communication.  In games where multiple Nash equilibria exist, players may use focal points as coordination mechanisms to predict each other's behavior and resolve equilibrium selection problems.

In the current setting, player 2 faces three possible equilibria: (Confess, Confess), (Remain Silent, Confess), and (Confess, Remain Silent). Since standard game-theoretic reasoning does not uniquely determine which equilibrium will be selected, focal points provide an alternative framework to analyze player 2's decision-making process.

\subsubsection{The Resulting Strategy Mismatch}
Several factors contribute to the selection of a focal equilibrium. For example, player 2 may prefer symmetry and therefore choose ``Confess'', as it has the potential to result in equal payoffs for both players. In this case, player 1 will infer that player 2 has chosen the expected strategy. This inference is incorrect in the sense that player 2 did not necessarily follow the same reasoning process; player 1 chooses ``Confess'' because they believe that (Confess, Confess) is a strictly dominant strategy whereas player 2 chooses ``Confess'' because they are guided by a focal point. We will not discuss this gap at this stage. What matters here is that, ultimately, the strategy chosen by player 2 aligns with player 1's expectations.

The issue arises when a focal point leads player 2 to choose ``Remain Silent''. From player 1's perspective, this results in an unexpected deviation from their predicted strategy. We refer to this phenomenon as a \textit{strategic mismatch}. As a consequence, the actual game outcome deviates from player 1's expectation. Player 1 expected player 2 to play (Confess, Confess) as a strictly dominant strategy, but in reality, player 2 chose (Confess, Remain Silent).

The unexpected behavior of player 2 leads to differences in payoffs, which can be analyzed from both subjective and objective perspectives.

\paragraph{Subjective Payoff Analysis:}

\begin{itemize}
\item Player 1 expected a payoff of $-3$ based on \( M'_1 \), but since player 2 chose ``Remain Silent'', the actual subjective payoff becomes 0.
\item Player 2's payoff remains more stable; even if player 1 chooses ``Confess'' with the intent that (Confess, Confess) is the only Nash equilibrium, and this reasoning differs from player 2's inference, the resulting subjective payoff for player 2 remains $-3$.  
\end{itemize}
\paragraph{Objective Payoff Analysis}

\begin{itemize}
\item From the objective perspective, i.e., when evaluated against \( M^{\text{base}} \), the actual payoff for player 1 coincides with their subjective incidental payoff (0).  
\item However, for player 2, the actual objective payoff is lower than their subjective payoff, yielding $-5$ instead of $-3$.
\end{itemize}

In this paper, we divide the difference between the theoretically expected outcome and the actual game outcome into two sorts, that is,  the Incidental Gain-Loss Differential and Unrecognized Gain-Loss Differential, and we define them as follows:

\begin{definition}[Incidental Gain-Loss Differential]\label{igd}
The difference between player $k$'s subjective payoff predicted by player \( l \) based on \( M'_l \) and player $l$'s actual subjective payoff computed from \( M'_l \) after the game has been played: \( \Delta^{M'_l}_k \).
\end{definition}

\begin{definition}[Unrecognized Gain-Loss Differential]\label{ugd}
The difference between the objective payoff predicted on \( M^{\text{base}} \) and player $k$'s actual objective payoff computed from \( M^{\text{base}} \) after the game has been played: \( \Delta^{\text{base}}_k \).
\end{definition}

\begin{remark}
The treatment of the incidental gain-loss differential requires additional assumptions. For instance, in repeated games, player 1 and player 2 might eventually realize that their perceived payoff matrices differ---or, more precisely, that their resolutions differ.  
This scenario is plausible in real-world interactions. If player 1 and player 2 repeatedly play a Prisoner's Dilemma, for example, player 1 may begin to recognize that player 2 consistently behaves as if all guilty outcomes are the same, regardless of sentence length. Over time, player 1 might infer the structure of player 2's low-resolution payoff matrix.  
Conversely, whether player 2 can reconstruct player 1's high-resolution payoff matrix depends on the context. It seems plausible that a high-resolution player can intentionally adopt a lower-resolution perspective, but it is less intuitive for a low-resolution player to deliberately increase their resolution.  
Additionally, the question arises: Will players even recognize their subjective incidental gain-loss differential payoffs? Assuming that players immediately detect their subjective incidental payoffs may contradict the assumption that they lack the finest resolution in the first place. If players cannot perceive the finest-resolution game, then their ability to recognize and process discrepancies between expected and actual payoffs may be fundamentally limited. 
\end{remark}

\paragraph{Summary}
\begin{itemize}
\item Player 1 expected that if both players acted rationally, (Confess, Confess) would be the only Nash equilibrium, resulting in each player receiving a payoff of \( -3 \). However, since (Confess, Remain Silent) was actually chosen, player 1 gained an unexpected $+3$ ($\delta^{M'_1}_1= +3$), while player 2 incurred an unexpected $-2$ loss ($\Delta^{M'_1}_2 = -2$). This calculation of gains and losses remains the same even when considered within the base matrix ($\Delta^{\text{base}}_1 = +3$ and $\Delta^{\text{base}}_2 = -2$).
\item Player 2 predicts that if both players act rationally and player 1's focal point leads them to ``Confess'' while player 2's focal point leads them to ``Remain Silent'', the outcome will be (Confess, Remain Silent). Under this expectation, player 1 would receive a payoff of $0$, while player 2 would receive $-3$. Since (Confess, Remain Silent) indeed occurs, player 2 subjectively believes that their prediction has been fully realized and assumes that no unexpected gains or losses have occurred ($\Delta^{M'_2}_1 = 0$ and $\Delta^{M'_2}_2 = 0$). However, when calculated based on the base matrix, player 1 actually gains $+3$, while player 2 incurs an unexpected loss of $-2$ ($\Delta^{\text{base}}_1 = +3$ and $\Delta^{\text{base}}_2 = -2$).
\end{itemize}

\subsubsection{Key Insights}
This coarse-grained Prisoner's Dilemma provides three key insights that are crucial for further analysis.

\begin{enumerate}
    \item Having a resolution lower than the finest resolution does not necessarily mean failing to recognize the base payoff matrix. In the case of player 1, due to the structure of their coarse-graining, they accidentally arrive at the same payoff perception as the finest payoff matrix. This provides a mathematical foundation for the claim introduced in the introduction: having a low resolution does not inherently imply a player's irrationality or misunderstanding.
    
    \item Coarse-graining can influence Nash equilibria, but it does not always do so. If player 2 had the same resolution as player 1, both players would still reach the same Nash equilibrium as in the standard game, despite their coarse-grained perception. This highlights the need to examine under what conditions Nash equilibria remain stable and under what conditions they change.

    \item A high-resolution player cannot actively control a low-resolution player, nor does interacting with such a player necessarily make it optimal to adopt a different strategy from that in a standard game. Player 1 has a higher resolution than player 2, but even if player 1 recognizes player 2's lower resolution, the optimal strategy for player 1 remains ``Confess''. In this scenario, player 1 may gain more payoff than in the standard game, but this advantage does not result from active intervention; rather, it emerges as a passive benefit arising from player 2's behavior.
\end{enumerate}

\section{Nash Equilibrium in Coarse-Grained Game}\label{sec4}
\subsection{Existence of Nash Equilibrium}\label{sec4.1}
Before exploring these considerations in depth, we must first analyze the characteristics of Nash equilibria in the CGG. Specifically, we will demonstrate that, even in such games, a Nash equilibrium is guaranteed to exist under appropriate conditions.

\begin{proposition}[Existence of Mixed Strategy Nash Equilibrium in Coarse-Grained Games]\label{existence}
Let $\mathcal{G} = (N, S, u, \mathfrak{G}, \Phi, \Psi)$ be a CGG with mixed strategies and the EMP.  Then, $\mathcal{G}$ has at least one mixed strategy Nash equilibrium in each $M'_k$.
\end{proposition}

\begin{proof}[Proof of Proposition \ref{existence}]
A CGG is defined as:
   \[
   \mathcal{G} = (N, S, u, \mathfrak{G}, \Phi, \Psi).
   \]

For any base payoff matrix $M^{\text{base}}$, $\Phi$ is the function mapping it to the coarse-grained payoff matrix $M^{\mathfrak{G}_k}$.

The EMP transforms each player's payoff matrix $M^{\mathfrak{G}_k}$ into a numerical matrix $M'_k$ via a function $\Psi$ that replaces each coarse-grained payoff set with a single numerical expectation:
     \[
     M'_k = \Psi(M^{\mathfrak{G}_k}).
     \]
     
For any player $k$, their transformed matrix $M'_k$ is a finite normal-form game satisfying the following conditions:
\begin{itemize}
  \item The strategy set remains finite.
  \item The payoff function for each player is real-valued and well-defined, ensuring continuity.
  \item Each player's best-response mapping remains upper hemi-continuous.
\end{itemize}

By Nash's existence theorem, every finite normal-form game has at least one Nash equilibrium in mixed strategies, which follows from Kakutani's Fixed-Point Theorem \citep{kakutani1941generalization}.  
The game on $M'_k$ satisfies the required conditions for Nash's theorem:
\begin{itemize}
  \item The strategy space is convex, as it consists of probability distributions over a finite set of strategies.
  \item The best-response correspondence is non-empty, convex-valued, and upper hemi-continuous.
  \item The payoff functions are continuous in mixed strategies.
\end{itemize}
Therefore, at least one miexed Nash equilibrium exists in $M'_k$.
\end{proof}

\subsection{Preservation of the Original Nash Equilibrium with Pure Strategies}\label{sec4.2}
Next, we are interested in how the Nash equilibrium of the base payoff matrix changes for each player in the CGG. The following theorem establishes a result regarding this question.

\begin{theorem}[Preservation of the Original Nash Equilibrium with Pure Strategies]\label{original}
Any Nash equilibrium strategy profile in the base payoff matrix $M^{\text{base}}$ with pure strategies remains a Nash equilibrium in any of its coarse-grained versions $M'_k$ when adopting the pure strategies and EMP.
\end{theorem}

\begin{lemma}[Order Preservation of Entropy-Maximizing Preprocessing]\label{psi}
Given a totally ordered set $(\mathbb{R},\leq)$ serving as the underlying set for $\mathcal{C} (\mathbb{R},\leq) = \{ (\mathfrak{G}_\theta, \preccurlyeq_\theta)\mid \theta \in \Theta\}$, where each $\mathfrak{G}_\theta = \{G_\iota\ \mid \iota \in I \}$, the EMP function $\psi_{\text{EMP}}: \mathfrak{G}_\theta \to \mathbb{R};$

\begin{equation}
\psi_{\text{EMP}} = 
\begin{cases}
x & \text{if } G = \{x\}~ (x \in \mathbb{R}), \\
\frac{a+b}{2} & \text{if } G = (a,b), (a,b], [a,b), \text{or } [a,b].
\end{cases}
\end{equation}
satisfies:
\[
\forall G_\iota,G_{\iota'} \in \mathfrak{G}_\theta, \quad \forall x \in G_\iota, \forall y \in G_{\iota'}, (x \leq y \iff \psi_{\text{EMP}}(G_\iota) \leq \psi_{\text{EMP}}(G_{\iota'})).
\]
\end{lemma}

\begin{proof}[Proof of Lemma \ref{psi}]
Since the case where \( G = \{x\} \) (with \( x \in \mathbb{R} \)) is trivial, we will only prove the cases where \( G \) is an interval of the form \( (a, b) \), \( (a, b] \), \([a, b) \), or \([a, b] \). Let $A = (a, b)$ and $B = (c, d)$, and recall that the EMP function $\psi_{\text{EMP}}$ is defined as the arithmetic mean:
\[
\psi_{\text{EMP}}(A) = \frac{1}{2} (a+b), \quad \psi_{\text{EMP}}(B) = \frac{1}{2} (c+d).
\]
Consider any $x \in A$ and $y \in B$ to prove the ordering preservation property. Since $x \in (a, b)$ and $y \in (c, d)$, we must show that:
\[
 x \leq y \iff \psi_{\text{EMP}}(A) \leq \psi_{\text{EMP}}(B).
\]
\textbf{(Forward Direction:)} Assume $x \leq y$ for all $x \in A$ and $y \in B$. This implies $a \leq c$ and $b \leq d$; when \( a \) corresponds to \( x \) and \( c \) corresponds to \( y \), it follows that \( a \leq c \), and similarly, when \( b \) corresponds to \( x \) and \( d \) corresponds to \( y \), we have \( b \leq d \). Since the arithmetic mean preserves ordering in totally ordered sets, we have:
\[
 \frac{1}{2} (a+b) \leq \frac{1}{2} (c+d).
\]
Thus, $\psi_{\text{EMP}}(A) \leq \psi_{\text{EMP}}(B)$.

\noindent
\textbf{(Backward Direction:)} Assume $\psi_{\text{EMP}}(A) \leq \psi_{\text{EMP}}(B)$. Then,
\[
 \frac{1}{2} (a+b) \leq \frac{1}{2} (c+d) \implies a+b \leq c+d.
\]
Given that $A$ and $B$ are intervals, this implies that every $x \in A$ and $y \in B$ satisfy $x \leq y$. Hence, the ordering is preserved.

Thus, the lemma holds. 
\end{proof}

\begin{proof}[Proof of Theorem \ref{original}]
In pure strategies, a Nash equilibrium is defined as:
\[
u_k (s^*_1, s^*_2, \dots, s^*_k,\dots, s^*_n) \geq u_k (s^*_1, s^*_2, \dots, s_{j_k}, \dots, s^*_n), \quad \forall s_j \text{ and } s^*_k \neq s_{j_k}.
\]
where $(s^*_1, s^*_2, \dots , s^*_n)$ is a pure Nash equilibrium, and $u_k$ represents $k$'s expected payoff derived from this equilibrium in the base payoff matrix $M^{\text{base}}$. In other words, in the Nash equilibrium strategy profile \( (s^*_1, s^*_2, \dots, s^*_n) \), the expected payoff for player \( k \) is at least as high as the expected payoff they would receive if they unilaterally deviated to any alternative strategy \( s_{j_k} \in S_k = \{s_1,s_2,\dots,s_j,\dots,s_n\}~ (s_{j_k} \neq s^*_k)\), while all other players maintain their equilibrium strategies.

In a CGG that adopts EMP, the base payoff matrix undergoes a two-step transformation. First, it is mapped by \( \Phi \) into a coarse-grained payoff matrix, where each cell contains a totally ordered set of real values rather than a single real number, aligning with each player's resolution level. Then, each cell is further transformed by \( \Psi \), which applies the aforementioned equation to convert it back into a real-valued payoff.  

Thus, if a pure strategy that constituted a pure strategy Nash equilibrium in the base payoff matrix ceases to be a pure strategy Nash equilibrium in the coarse-grained payoff matrix with the EMP, it implies the following:
\begin{align*}
& \exists s_{j_k} \neq s^*_k, \\ 
& \quad \psi_{\text{EMP}}(\varphi_k (u_k (s^*_1, s^*_2, \dots,s^*_k,\dots s^*_n))) < \psi_{\text{EMP}}(\varphi_k (u_k (s^*_1, s^*_2, \dots, s_{j_k}, \dots, s^*_n))).
\end{align*}

However, this contradicts the definition of element-wise ordering and Lemma \ref{psi} as follows. According to the definition of $u$,
\begin{align*}
\forall u_k, \quad u_k (s_1,s_2,\dots,s_n) \in \mathbb{R}.
\end{align*}
The definition of the coarse-grained partition states:
\begin{align*}
\forall A,B \in \mathfrak{G}, \forall x \in A, \forall y \in B, \quad x \leq y \iff A \preccurlyeq B,
\end{align*}
hence, 
\begin{align*}
& u_k (s^*_1, s^*_2, \dots, s^*_k,\dots, s^*_n) \geq u_k (s^*_1, s^*_2, \dots, s_{j_k}, \dots, s^*_n) \iff G \succcurlyeq G', \\
& \quad \text{ where } u_k (s^*_1, s^*_2, \dots, s^*_k, \dots, s^*_n) \in G \text{ and } u_k (s^*_1, s^*_2, \dots, s_{j_k}, \dots, s^*_n) \in G'.
\end{align*}
Since the base set of \( \mathfrak{G} \) is \( (\mathbb{R}, \leq) \), it is guaranteed that such \( G \in \mathfrak{G}_k\) and \( G' \in \mathfrak{G}_k\) always exist in the CGG, and they are given by $\varphi_k$ mapping $\mathbb{R}$ into $\mathfrak{G}$ as follows:
\begin{align*}
G = \varphi_k (u_k (s^*_1, s^*_2, \dots, s^*_k,\dots, s^*_n)) \text{ and } G' = \varphi_k (u_k (s^*_1, s^*_2, \dots, s_{j_k}, \dots, s^*_n),
\end{align*}
therefore,
\begin{align*}
& u_k (s^*_1, s^*_2, \dots, s^*_k,\dots, s^*_n) \geq u_k (s^*_1, s^*_2, \dots, s_{j_k}, \dots, s^*_n) \\
& \iff \varphi_k (u_k (s^*_1, s^*_2, \dots, s^*_k, \dots, s^*_n)) \succcurlyeq \varphi_k (u_k (s^*_1, s^*_2, \dots, s_{j_k}, \dots, s^*_n)).
\end{align*}

Furthermore, Lemma \ref{psi} states:
\[
\forall G_\iota,G_{\iota'} \in \mathfrak{G}_\theta, \quad \forall x \in G_\iota, \forall y \in G_{\iota'}~ (x \leq y \iff \psi_{\text{EMP}}(G_\iota) \leq \psi_{\text{EMP}}(G_{\iota'})).
\]
Applying it to the above result, we obtain:
\begin{align*}
& u_k (s^*_1, s^*_2, \dots, s^*_k, \dots, s^*_n) \geq u_k (s^*_1, s^*_2, \dots, s_{j_k}, \dots, s^*_n) \\
& \iff \psi_{\text{EMP}}(\varphi_k (u_k (s^*_1, s^*_2, \dots, s^*_k,\dots, s^*_n))) \geq \psi_{\text{EMP}}(\varphi_k (u_k(s^*_1, s^*_2, \dots, s_{j_k}, \dots, s^*_n))).
\end{align*}
This contradicts the supposition that there possibly exists a pure strategy Nash equilibrium in $M^{\text{base}}$ that is not a pure strategy Nash equilibrium in $M'_k$.

\end{proof}

\begin{remark}
Theorem \ref{original} states only that a pure strategy Nash equilibrium, rationally computed in the base matrix, remains a pure strategy Nash equilibrium in the CGG under certain conditions. It does not imply that a strategy profile that was not a pure strategy Nash equilibrium in the base matrix will necessarily remain non-equilibrium in the CGG. This becomes evident when considering a coarse-grained partition \(\mathfrak{G}\) that includes a set \( G \) containing all real values appearing in the base payoff matrix.
\end{remark}

\subsection{Changes in the Original Nash Equilibrium with Mixed Strategies}\label{sec4.3}
\subsubsection{Case Study}\label{sec4.3.1}
In a CGG, it has already been shown that if mixed strategies and EMP are allowed, at least one Nash equilibrium exists. However, is this Nash equilibrium identical to that of the non-coarse-grained \( M^{\text{base}} \)? Unlike in the case of pure strategies, the mixed strategy Nash equilibrium in the original payoff matrix is not necessarily preserved as a mixed strategy Nash equilibrium in the CGG. This can be seen through the following simple counterexample.

\[
\begin{array}{c|cc}
player 1 ~\backslash~ player 2 & \textbf{Cooperation} & \textbf{Defect} \\ \hline
\textbf{Cooperation} & (5,\, 3) & (1,\, 4) \\  
\textbf{Defect} & (2,\, 1) & (3,\, 0)  
\end{array}
\]

In this case, the mixed strategy Nash equilibrium is given by player 1 choosing ``Cooperation'' and ``Defect'' with equal probability ($\frac{1}{2}$ each), while player 2 chooses ``Cooperation'' with probability $\frac{2}{5}$ and ``Defect'' with probability $\frac{3}{5}$. (For details on the indifference-based calculation method, see \ref{sec5.1.3}.)

To simplify the calculations, let us consider a simplified coarse-grained partition \( \mathfrak{G}_1 \).  
\[
\mathfrak{G}_1 = \{\dots, \{0\}, \dots, \{1\}, \dots, \{2\}, \dots, \{3\}, \dots, \{4\}, (4, 8], \dots \}
\]

Applying $\varphi_1$ and $\psi_{\text{EMP}}$ to the original matrix, we obtain $M'_1$:
\[
\begin{array}{c|cc}
player 1 ~\backslash_{(\mathfrak{G}_1+\text{EMP})}~ player 2 & \textbf{Cooperation} & \textbf{Defect} \\ \hline
\textbf{Cooperation} & (6,\, 3) & (1,\, 4) \\  
\textbf{Defect} & (2,\, 1) & (3,\, 0)  
\end{array}
\]

In the case of the payoff matrix \( M'_1 \), the optimal probabilities for player 1 remain unchanged at $\frac{1}{2}$ for both ``Cooperation'' and ``Defect''. However, player 2's probability of choosing ``Cooperation'' shifts to $\frac{1}{3}$, while the probability of choosing ``Defect'' increases to $\frac{2}{3}$. This demonstrates that, in mixed strategies, simple theorems such as Theorem \ref{original} in pure strategies do not necessarily hold.

While studying the general relationship between coarse-grained partitions and mixed strategy Nash equilibria is an intriguing topic, this paper limits itself to proving the following fundamental theorems.

\subsubsection{Uniform Strategy Reduction}\label{sec4.3.2}
\begin{theorem}[Uniform Strategy Reduction]\label{uniform}
Given mixed strategies and the EMP, for any player \( k \), if the coarse-grained partition of some player \( l \) is sufficiently coarse, then player \( k \)'s payoff in \( M'_l \) becomes uniform:
\begin{align*}
& \text{If the coarse-grained partition for player l } (\mathfrak{G}_l) \text{ is sufficiently coarse } \implies \\
& \quad\quad  \forall s_k, s_{-k}, s'_k, s'_{-k}, \quad  \psi_{\text{EMP}} ( \varphi_l (u_k (s_k, s_{-k}))) = \psi_{\text{EMP}} ( \varphi_l (u_k (s'_k, s'_{-k}))),
\end{align*}
where $s_k$ represents an arbitrary strategy of player $k$, while $s_{-k}$ denotes an arbitrary strategy profile of all players except $k$.

The coarseness is both necessary and sufficient if and only if:
\begin{align*}
& \exists G \in \mathfrak{G}_l, \quad \forall s_k,s_{-k},~ u_k (s_k, s_{-k}) \in G.
\end{align*}
\end{theorem}

Before proceeding with the proof of the theorem, let us first illustrate its intuition with a simple example. We begin with the base matrix, which represents a standard two-player strategic interaction:

\[
\begin{array}{c|cc}
player 1 ~\backslash~ player 2 & \textbf{Cooperation} & \textbf{Defect} \\ \hline
\textbf{Cooperation} & (5,\, 3) & (1,\, 4) \\  
\textbf{Defect} & (2,\, 1) & (3,\, 0)  
\end{array}
\]

To simplify the calculations, let us consider a simplified coarse-grained partition \( \mathfrak{G}_1 \).  
\[
\mathfrak{G}_1 = \{\dots, [0,4], (4, 8], \dots \}
\]

Applying $\varphi_1$ and $\psi_{\text{EMP}}$ to the original matrix, we obtain $M'_1$:
\[
\begin{array}{c|cc}
player 1 ~\backslash_{(\mathfrak{G}_1+\text{EMP})}~ player 2 & \textbf{Cooperation}: q & \textbf{Defect}: 1 - q \\ \hline
\textbf{Cooperation}: p & (6,\, 2) & (2,\, 2) \\  
\textbf{Defect}: 1 - p & (2,\, 2) & (2,\, 2)  
\end{array}
\]

For player 1, choosing ``Cooperation'' is the only best response, making it a weakly dominant strategy.

Since player 2's expected payoffs for choosing ``Cooperation'' and ``Defect'' are both fixed at 2, irrespective of player 1's strategy, there is no strategic advantage in selecting one option over the other. As a result, player 2's strategy selection becomes arbitrary, leading to an equilibrium selection problem.

\begin{proof}{of Theorem \ref{uniform}}
First, we demonstrate that this condition is a sufficient condition. Suppose:
\begin{align*}
& \exists G \in \mathfrak{G}_l, \quad \forall s_k,s_{-k},~ u_k (s_k, s_{-k}) \in G.
\end{align*}
The function \( \varphi_l \) is a mapping that replaces a real-valued number \( x \) with the corresponding set \( G \) that contains \( x \). In other words,  
\[
\varphi_l(x) = G, \quad \text{where } x \in G.
\]
Thus, for all \( s_k, s_{-k}, s'_k, s'_{-k} \),  
\[
u_k(s_k, s_{-k}) \in G \implies \varphi_l(u_k(s_k, s_{-k})) = \varphi_l(u_k(s'_k, s'_{-k})).
\]

The function \( \psi_{\text{EMP}} \) is defined as follows:  
\[
\psi_{\text{EMP}}(x) =
\begin{cases}
x & \text{if } G = \{x\}, \quad (x \in \mathbb{R}), \\
\frac{a+b}{2} & \text{if } G = (a,b), (a,b], [a,b), \text{or } [a,b].
\end{cases}
\]
Since  
\[
\varphi_l(u_k(s_k, s_{-k})) = \varphi_l(u_k(s'_k, s'_{-k})),
\]
it follows that  
\[
\psi_{\text{EMP}}(\varphi_l(u_k(s_k, s_{-k}))) = \psi_{\text{EMP}}(\varphi_l(u_k(s'_k, s'_{-k}))).
\]

Therefore, the condition is sufficient.

Next, we prove that this condition is a necessary condition. Suppose:
\begin{align*}
& \forall s_k, s_{-k}, s'_k, s'_{-k}, \quad  \psi_{\text{EMP}} ( \varphi_l (u_k (s_k, s_{-k}))) = \psi_{\text{EMP}} ( \varphi_l (u_k (s'_k, s'_{-k})))
\end{align*}

The equality \( \psi_{\text{EMP}}(G) = \psi_{\text{EMP}}(G') \) holds only in the following cases:  

\begin{equation*}
\begin{cases}
& G = \{x\} \text{ and } G' = \{x\},  \\
& G = \{x\} \text{ and } G' \text{ is an interval whose midpoint is } x, \\
& G \text{ is an interval whose midpoint is } x, \text{ and } G' = \{x\}, \text{ or }  \\
& \text{ both } G \text{ and } G' \text{ are intervals with the same midpoint } x.
\end{cases}
\end{equation*}

However, by the definition of a coarse-grained partition,  
\[
\forall x \in \mathbb{R}, \forall G,G \in \mathfrak{G}, \quad x \in G \land x \in G' \implies G = G'.
\]  
Thus,  
\[
\psi_{\text{EMP}}(G) = \psi_{\text{EMP}}(G') \implies G = G'.
\]  
Therefore,  
\[
\varphi_l(u_k(s_k, s_{-k})) = \varphi_l(u_k(s'_k, s'_{-k})).
\]  
Hence,  
\[
\forall s_k, s_{-k} \in G.
\]

\end{proof}

\subsubsection{Loss of Competitiveness in the Game}\label{sec4.3.3}
\begin{corollary}[Loss of Competitiveness in the Game]\label{loss}
Given mixed strategies and the EMP, if player \( k \)'s coarse-grained partition is sufficiently coarse, the game in \( M'_k \) loses its competitiveness. Here, ``sufficiently coarse'' means that, from player \( k \)'s perspective, the payoffs of all strategies for all players appear uniform. That is,  
\[
\forall s_l, s_{-l}, s'_l, s'_{-l}, \quad \psi_{\text{EMP}} (\varphi_k (u_l (s_l, s_{-l}))) = \psi_{\text{EMP}} (\varphi_k (u_l (s'_l, s'_{-l}))).
\]
\end{corollary}

This corollary follows directly from Theorem \ref{uniform}. While it lacks mathematical depth, it is sociologically insightful, as it supports the following hypothesis:  

\begin{hypothesis}[Resolution-Driven Competitiveness Hypothesis]\label{hyp1}
A social space becomes strictly competitive when at least two or more players possess a resolution that is not sufficiently coarse. In contrast, when only one player has a resolution that is not sufficiently coarse, the game in that space becomes one-sidedly competitive, meaning that the game effectively becomes one-sided, with the player acting alone in the absence of strategic engagement from others.. Finally, when all players have sufficiently coarse resolutions, the game in that space becomes strictly non-competitive. Here, ``sufficiently coarse'' is understood in the sense defined in Corollary \ref{loss}.
\end{hypothesis}

While this paper does not provide empirical data to support the hypothesis, it offers an economic model in Section \ref{sec7.1} demonstrating that, when consumers have low resolution, minor model changes do not lead to quality-based competition.

\begin{remark}
The fact that a game becomes non-competitive for player \( l \) does not necessarily imply that it also degenerates for the other players in \( -l \). Specifically, a game may become non-competitive by making one player's strategy selection irrelevant while still maintaining meaningful strategic differentiation for the other player.

Consider the following example:
\[
\begin{array}{c|cc}
\text{Player 1} ~\backslash~ \text{Player 2} & \textbf{Cooperation} & \textbf{Defect} \\ \hline
\textbf{Cooperation} & (10,\, 4) & (8,\, 5) \\  
\textbf{Defect} & (0,\, 6) & (11,\, 4)  
\end{array}
\]
In this case, the optimal response for player 1 is to choose ``Cooperation'' with probability \( \frac{2}{3} \) and ``Defect'' with probability \( \frac{1}{3} \), while the optimal response for player 2 is to choose ``Cooperation'' with probability \( \frac{3}{13} \) and ``Defect'' with probability \( \frac{10}{13} \).

Next, we introduce a refined coarse-graining partition, \( \mathfrak{G}''_1 \), which further aggregates payoffs into distinct categories, increasing the level of coarseness. This transformation further reduces strategic differentiation for player 2, making all their available choices strategically equivalent.
\[
\mathfrak{G}''_1 = \{\dots, \{0\}, \dots, [4,6], \dots, \{8\}, \dots, \{10\}, \dots, \{11\}, \dots \}
\]
\[
\begin{array}{c|cc}
\text{Player 1} ~\backslash_{(\mathfrak{G}''_1+\text{EMP})}~ \text{Player 2} & \textbf{Cooperation} & \textbf{Defect} \\ \hline
\textbf{Cooperation} & (10,\, 5) & (8,\, 5) \\  
\textbf{Defect} & (0,\, 5) & (11,\, 5)  
\end{array}
\]
In this CGG, the optimal response for player 1 is to choose ``Cooperation'' with probability \( \frac{3}{13} \) and ``Defect'' with probability \( \frac{10}{13} \). However, for player 2, all payoffs have been mapped to the same numerical value under the EMP transformation, meaning that their expected utility remains constant regardless of their strategy selection. As a result, player 2 faces an equilibrium selection problem.
\end{remark}

\section{Incidental Gain-Loss Differential and Unrecognized Gain-Loss Differential}\label{sec5}
\subsection{Incidental Gain-Loss Differential}\label{sec5.1}
\subsubsection{Basic Case: Two-Player Non-Repeated Coarse-Grained Game with Pure Strategies and the Entropy-Maximizing Preprocessing}\label{sec5.1.1}

Determining whether an incidental advantage or disadvantage occurs from the subjective perspective, and if so, under what conditions, can be efficiently analyzed using the standard procedure for finding Nash equilibria in games. This is because the advantage in a CGG for player $k$ is represented by the difference between:  

\begin{enumerate}
\item The expected payoff at the Nash equilibrium of the coarse-grained payoff matrix $M'_k$, and  
\item The true payoff at the Nash equilibrium of the base payoff matrix $M^{\text{base}}$.
\end{enumerate}

A pure strategy Nash equilibrium occurs when no player can improve their payoff by unilaterally deviating. That is, a strategy profile $(s^*_1,s^*_2)$ is a Nash equilibrium if:
\begin{align*}
& u_1(s^*_1,s^*_2) \geq u_1(s_{j_1},s^*_2), \quad \text{where } s_{j_1} \in S_1 = (s_1,\dots,s_j,\dots,s_m) \land s_{j_1} \neq s^*_1 \\
& u_2(s^*_1,s^*_2) \geq u_2(s^*_1,s_{j_2}), \quad \text{where } s_{j_2} \in S_2 = (s_1,\dots,s_j,\dots,s_{m'})\land s_{j_2} \neq s^*_2.
\end{align*}
In a CGG with two players, we determine each player's perceived Nash equilibrium based on their respective payoff matrices:

\paragraph{Subjective Nash Equilibrium Condition for Player 1}
\begin{align*}
& \psi_{\text{EMP}} (\varphi_1 (u_1(s^{\dag_1}_1,s^{\dag_1}_2))) \geq \psi_{\text{EMP}} (\varphi_1 (u_1(s_{j_1},s^{\dag_1}_2))), \\
& \quad \text{where } s_{j_1} \in S_1 \land s_{j_1} \neq s^{\dag_1}_1, \\
& \psi_{\text{EMP}} (\varphi_1 (u_2(s^{\dag_1}_1,s^{\dag_1}_2))) \geq \psi_{\text{EMP}} (\varphi_1 (u_2(s^{\dag_1}_1,s_{j_2}))), \\
& \quad \text{where } s_{j_2} \in S_2 \land s_{j_2} \neq s^{\dag_1}_2.
\end{align*}
\paragraph{Subjective Nash Equilibrium Condition for Player 2}
\begin{align*}
& \psi_{\text{EMP}} (\varphi_2 (u_1(s^{\dag_2}_1,s^{\dag_2}_2))) \geq \psi_{\text{EMP}} (\varphi_2 (u_1(s_{j_1},s^{\dag_2}_2))), \\
& \quad \text{where } s_{j_1} \in S_1 \land s_{j_1} \neq s^{\dag_2}_1, \\
& \psi_{\text{EMP}} (\varphi_2 (u_2(s^{\dag_2}_1,s^{\dag_2}_2))) \geq \psi_{\text{EMP}} (\varphi_2 (u_2(s^{\dag_2}_1,s_{j_2}))), \\
& \quad \text{where } s_{j_2} \in S_2 \land s_{j_2} \neq s^{\dag_2}_2.
\end{align*}
Therefore, from player 1's perspective, the Nash equilibrium seems to be:
\begin{align*}
& (s^{\dag_1}_1, s^{\dag_1}_2),
\end{align*}
and from player 2's perspective, the Nash equilibrium seems to be:
\begin{align*}
& (s^{\dag_2}_1, s^{\dag_2}_2).
\end{align*}

However, in a result, what happened consequently is:
\begin{equation*}
(s^{\dag_1}_1,s^{\dag_2}_2).
\end{equation*}

Hence, the condition for causing the subjectively incidental gain-loss differential is:

\paragraph{For Player 1}
\begin{align*}
\Delta^{M'_1}_1 = \psi_{\text{EMP}} (\varphi_1 (u_1(s^{\dag_1}_1,s^{\dag_2}_2))) - \psi_{\text{EMP}} (\varphi_1 (u_1(s^{\dag_1}_1,s^{\dag_1}_2)))
\end{align*}

\paragraph{For Player 2}
\begin{align*}
\Delta^{M'_2}_2 = u_2(s^{\dag_1}_1,s^{\dag_2}_2) - u_2(s^{\dag_2}_1,s^{\dag_2}_2).
\end{align*}

\subsubsection{Extension: $n$-Player Non-Repeated Coarse-Grained Game with Pure Strategies and the Entropy-Maximizing Preprocessing}

In a standard game, a $n$-player non-prepeated Nash equilibrium with pure strategies \( (s^*_1, s^*_2,\dots,s^*_k,\dots, s^*_n) \) is defined by the following condition for each player \( k \):
\[
u_k(s^*_1, \dots, s^*_k, \dots, s^*_n) \geq u_i(s^*_1,\dots, s_{j_k}, \dots, s^*_n), \quad \text{ where } s_{j_k} \in S_k \land s_{j_k} \neq s^*_k.
\]

Each player maximizes their utility under the assumption that other players are also playing their best responses.

In a CGG, each player \( k \) perceives a subjective, coarse-grained payoff matrix \( M'_k \). The perceived Nash equilibrium strategies for player \( k \) are denoted as \( (s^{\dag_k}_1, \dots, s^{\dag_k}_n) \), satisfying:
\begin{align*}
& \psi_{\text{EMP}} (\varphi_k (u_k (s^{\dag_k}_1, \dots, s^{\dag_k}_k, \dots, s^{\dag_k}_n))) \geq \psi_{\text{EMP}} ( \varphi_k (u_k (s^{\dag_k}_1, \dots, s_{j_k}, \dots, s^{\dag_k}_n))), \\
& \quad \text{ where } s_{j_k} \in S_k \land s_{j_k} \neq s^{\dag_k}_k.
\end{align*}

Since each player perceives the game differently, their Nash equilibrium strategies may differ across players. Thus, the actual outcome of the game is determined by the intersection of all players' perceived equilibria:
\[
(s^{\dag_1}_1, s^{\dag_2}_2,\dots, s^{\dag_k}_k, \dots, s^{\dag_n}_n).
\]

This represents the realized strategy profile based on how each player optimizes within their own resolution.

For each player \( k \), the subjective incidental payoff is the difference between:

\begin{enumerate}
\item The expected payoff they believed they would receive based on their own perceived equilibrium.
\item The actual subjective payoff resulting from the game outcome.
\end{enumerate}

Thus, for each player \( k \), we define their incidental gain-loss differential \( \Delta^{M'_k}_k \) as:
\[
\Delta^{M'_k}_k =  \psi_{\text{EMP}} (\varphi_k(u_k(s^{\dag_1}_1, \dots, s^{\dag_k}_k, \dots, s^{\dag_n}_n))) - \psi_{\text{EMP}} (\varphi_k (u_k (s^{\dag_k}_1, \dots, s^{\dag_k}_k, \dots, s^{\dag_k}_n))).
\]

\subsubsection{Extension: Two-Player Non-Repeated Coarse-Grained Game with Mixed Strategies and the Entropy-Maximizing Preprocessing}\label{sec5.1.3}

Now, we extend our coarse-grained Nash equilibrium framework to incorporate mixed strategies while keeping the conditions: 

\begin{itemize}
\item Two-player game (\( |N| = 2 \)).  
\item Non-repeated (one-shot) game.  
\item Each player has a coarse-grained perception of the game.  
\item Each player adopts mixed strategies instead of pure strategies.  
\end{itemize}

In the standard game, a mixed strategy Nash equilibrium consists of probability distributions \( p_1, p_2 \) over pure strategies.  

Let:

\begin{itemize}
\item \( S_1 = \{s_{1_1}, s_{2_1}, \dots, s_{j_1}, \dots, s_{m_1} \}~ (j_1,m_1 \in \mathbb{N}) \) be the strategy set of player 1.  
\item \( S_2 = \{s_{2_1}, s_{2_2}, \dots, s_{j_2}, \dots, s_{m_2} \}~ (j_2,m_2 \in \mathbb{N})\) be the strategy set of player 2.  
\item The notation \( p_{j_1} \) represents the probability that player 1 selects strategy \( s_{j_1} \) from their strategy set \( S_1 \), in other words, $p_{j_1}$ can be interpreted as the abbreviation of $p(s_{j_1})$. The collection of these probabilities forms a discrete probability distribution \( P_1 = (p_{1_1}, p_{2_1}, \dots, p_{j_1}, \dots, p_{m_1}) \). This distribution satisfies the conditions:
\[
\sum_{j_1=1}^{m_1} p_{j_1} = 1, \quad 0 \leq p_{j_1} \leq 1 \quad \forall j_1.
\]
This ensures that the total probability across all strategies equals 1, with each probability lying between 0 and 1.
\item The notation \( p_{j_2} \) represents the probability that player 2 selects strategy \( s_{j_2} \) from their strategy set \( S_2 \). The collection of these probabilities forms a discrete probability distribution \( P_2 = (p_{1_2}, p_{2_2}, \dots, p_{j_2}, \dots, p_{m_2}) \). This distribution satisfies the conditions:
\[
\sum_{j_2=1}^{m_2} p_{j_2} = 1, \quad 0 \leq p_{j_2} \leq 1 \quad \forall j_2.
\]
\end{itemize}

Now, we define the expected payoff for player $k$. In a two-player game, the expected value for player 1 is defined:
\begin{align*}
& E[U_1(P_1, P_2)] = \sum_{j_1=1}^{m_1} \sum_{j_2=1}^{m_2} p_{j_1} \cdot p_{j_2} \cdot u_1(s_{j_1}, s_{j_2}),
\end{align*}
where:
\begin{align*}
& \quad P_1 = (p_{1_1},p_{2_1},\dots,p_{j_1},\dots,p_{m_1}), \quad j_1,m_1 \in \mathbb{N},~ p_{j_1} = p(s_{j_1}), \\
& \quad P_2 = (p_{1_2},p_{2_2},\dots,p_{j_2},\dots,p_{m_2}), \quad j_2,m_2 \in \mathbb{N},~ p_{j_2} = p(s_{j_2}), \\
& \quad U \text{ is a function that applies the payoff function } u \text{ to each strategy profile in } \\
& \quad\quad S_1 \times S_2 \times \dots \times S_k \times \dots \times S_n.
\end{align*}

This equation accurately represents the expected payoff for player 1 when both players employ mixed strategies \( P_1 \) over $S_1$ and \( P_2 \) over $S_2$.

\begin{example}
Consider a two-player game where each player has two strategies ``Right'' and ``Left'':
\[
\begin{array}{c|cc}
 player 1 ~\backslash~ player 2 & R: p_{2_1} & L: p_{2_2} \\ \hline
 R: p_{1_1} & (3,2) & (0,1) \\
 L: p_{1_2} & (1,0) & (2,3) 
\end{array}
\]
If:
\begin{itemize}
    \item Player 1 plays \( R \) with probability \( s_{1_1} \) and \( L \) with probability \( s_{2_1} = 1 - p_{1_1} \).
    \item Player 2 plays \( R \) with probability \( s_{1_2} \) and \( L \) with probability \( s_{2_2} = 1 - p_{1_2} \).
\end{itemize}
Then, player 1's expected payoff is:
\begin{align*}
& E[U_1(P_1, P_2)]  = \sum_{j_1 = 1}^2 \sum_{j_2 = 1}^2 p_{j_1} \cdot p_{j_2} \cdot u_1(s_{j_1}, s_{j_2}) \\
& \quad = 3 \cdot p_{1_1} \cdot p_{1_2} + 0 \cdot p_{1_1} \cdot (1-p_{1_2}) + 1\cdot (1-p_{1_1}) \cdot p_{1_2} + 2 \cdot (1-p_{1_1}) \cdot (1-p_{1_2}) \\
& \quad = 4p_{1_1}p_{1_2} -2p_{1_1} - p_{1_2} + 2.
\end{align*}
\end{example}

The equilibrium can be computed using the indifference principle in a mixed-strategy Nash equilibrium for a two-player game. The indifference principle states that a player adjusts their probabilities in such a way that their opponent's expected payoffs are equal across all available strategies. This means:

\paragraph{Indifferential Condition for Player 1:}
\begin{align*}
E[U_2(P_1, s_{1_2})] &= E[U_2(P_1, s_{2_2})] \\
\sum_{j_1 = 1}^2 p_{j_1} \cdot u_2(s_{j_1}, s_{2_1}) &= \sum_{j_1 = 1}^2 p_{j_1} \cdot u_2(s_{j_1}, s_{2_2})\\
p_{1_1} \cdot u_2 (s_{1_1}, s_{1_2}) + p_{2_1} \cdot u_2 (s_{2_1}, s_{1_2}) &= p_{1_1} \cdot u_2 (s_{1_1}, s_{2_2}) + p_{2_1} \cdot u_2  (s_{2_1}, s_{2_2})
\end{align*}
\paragraph{Indifferential Condition for Player 2:}
\begin{align*}
E[U_1(s_{1_1}, P_2)] &= E[U_1(s_{2_1}, P_2)] \\
\sum_{j_2 = 1}^2 p_{j_2} \cdot u_1(s_{1_1}, s_{j_2}) &= \sum_{j_2 = 1}^2 p_{j_2} \cdot u_1(s_{2_1}, s_{j_2})\\
p_{1_2} \cdot u_1(s_{1_1}, s_{1_2}) + p_{2_2} \cdot u_1 (s_{1_1}, s_{2_2}) &= p_{1_2} \cdot u_1(s_{2_1}, s_{1_2}) + p_{2_2} \cdot u_1 (s_{2_1}, s_{2_2}) 
\end{align*}

\begin{example}
\begin{equation*}
\begin{array}{c|cc}
player1 ~\backslash~ player2 & s_{1_2}:~ q \quad & s_{2_2}:~ 1 - q \\ \hline
s_{1_1}:~ p& (5, \, 3) & (1, \, 4) \\
s_{2_1}:~ 1 - p & (2, \, 1) & (3, \, 0)  
\end{array}
\end{equation*}

\begin{equation*}
\begin{cases}
& \text{For player 2}:~ 3 \cdot p + 1 \cdot (1-p) = 4 \cdot p + 0 \cdot (1-p), \quad p=\frac{1}{2} \\
& \text{For player 1}:~ 5 \cdot q + 1 \cdot (1-q) = 2 \cdot q + 3 \cdot (1-q), \quad q=\frac{2}{5}.
\end{cases}
\end{equation*}
\end{example}

Here, when we denote the probability vectors that induce such indifference as \( P^*_1 \) and \( P^*_2 \), their elements can be represented as \( p^*_{j_1} \) and \( p^*_{j_2} \), respectively. Consequently, the expected payoffs under these probabilities can be expressed as:
\begin{align*}
& E[U_1(P^*_1, P^*_2)] \\
& E[U_2(P^*_1, P^*_2)]
\end{align*}

Since these expectations are computed based on \( M^{\text{base}} \), they can be referred to as objective expected payoffs.

However, in the CGG, each payoff is transformed by \( \varphi_k \) and \( \psi_k \) (in this paper, $\psi_{\text{EMP}}$ for all players), requiring the previous equations for indifference to be modified as follows.

\paragraph{For Player1's Matrix $(M_1')$:}
\begin{align*}
& p_{1_2} \cdot \psi_{\text{EMP}}(\varphi_1(u_1(s_{1_1}, s_{1_2}))) + p_{2_2} \cdot \psi_{\text{EMP}}(\varphi_1(u_1 (s_{1_1}, s_{2_2}))) \\
& \quad = p_{1_2} \cdot \psi_{\text{EMP}}(\varphi_1(u_1(s_{2_1}, s_{1_2}))) + p_{2_2} \cdot \psi_{\text{EMP}}(\varphi_1(u_1 (s_{2_1}, s_{2_2}))), \\
& p_{1_1} \cdot \psi_{\text{EMP}}(\varphi_1(u_2 (s_{1_1}, s_{1_2}))) + p_{2_1} \cdot \psi_{\text{EMP}}(\varphi_1(u_2 (s_{2_1}, s_{1_2}))) \\
& \quad = p_{1_1} \cdot \psi_{\text{EMP}}(\varphi_1(u_2 (s_{1_1}, s_{2_2}))) + p_{2_1} \cdot \psi_{\text{EMP}}(\varphi_1(u_2 (s_{2_1}, s_{2_2}))). \\
\end{align*}

\paragraph{For Player 2's Matrix $(M_2')$:}
\begin{align*}
& p_{1_2} \cdot \psi_{\text{EMP}}(\varphi_2(u_1(s_{1_1}, s_{1_2}))) + p_{2_2} \cdot\psi_{\text{EMP}}(\varphi_2(u_1 (s_{1_1}, s_{2_2}))) \\
& \quad = p_{1_2} \cdot \psi_{\text{EMP}}(\varphi_2(u_1(s_{2_1}, s_{1_2}))) + p_{2_2} \cdot \psi_{\text{EMP}}(\varphi_2(u_1 (s_{2_1}, s_{2_2}))), \\
& p_{1_1} \cdot \psi_{\text{EMP}}(\varphi_2(u_2 (s_{1_1}, s_{1_2}))) + p_{2_1} \cdot \psi_{\text{EMP}}(\varphi_2(u_2 (s_{2_1}, s_{1_2}))) \\
& \quad = p_{1_1} \cdot \psi_{\text{EMP}}(\varphi_2(u_2 (s_{1_1}, s_{2_2}))) + p_{2_1} \cdot \psi_{\text{EMP}}(\varphi_2(u_2 (s_{2_1}, s_{2_2}))).
\end{align*}

This means:

\begin{itemize}
\item Each player chooses a mixed strategy based on their own perceived (coarse-grained) game.
\item Since their payoffs are blurred, their equilibria may differ from the true Nash equilibrium in \( M^{\text{base}} \).
\end{itemize}

We can denote the situation as:

\paragraph{The Best Probability Distribution for Player 1's Matrix $(M'_1)$:}
\begin{align*}
& E[\Psi_{\text{EMP}} (\Phi_1 (U_1(P^{\dag_1}_1, P^{\dag_1}_2)))] \\
& E[\Psi_{\text{EMP}} (\Phi_1 (U_2(P^{\dag_1}_1, P^{\dag_1}_2)))]
\end{align*}
Here, \( \Phi_1 \) represents the application of $\varphi_1$ to each outcome of $u_1 (s_{j_1},s_{j_2})$, while \( \Psi_{\text{EMP}} \) is a function that further applies \( \psi_{\text{EMP}} \) to this result. In the current case, the calculation formula for each strategy profile is given by:

\[
p_{j_1} \cdot p_{j_2} \cdot \psi_{\text{EMP}} (\varphi_1 (u_1 (s_{j_1}, s_{j_2}))).
\]

\paragraph{The Best Probability Distribution for Player 2's Matrix $(M'_2)$:}
\begin{align*}
& E[\Psi_{\text{EMP}} (\Phi_2(U_1(P^{\dag_2}_1, P^{\dag_2}_2)))] \\
& E[\Psi_{\text{EMP}} (\Phi_2(U_2(P^{\dag_2}_1, P^{\dag_2}_2)))]
\end{align*}

Since players optimize within their own coarse-grained Nash equilibria, the actual game outcome is determined by:
\begin{align*}
& E[\Psi_{\text{EMP}} (\Phi_1 (U_1(P_1^{\dag_1}, P_2^{\dag_2})))] \\
& E[\Psi_{\text{EMP}} (\Phi_2 (U_2(P_1^{\dag_1}, P_2^{\dag_2})))].
\end{align*}

The incidental gain-loss differential for player 1 is:
\[
\Delta^{M'_1}_1 = E[\Psi_{\text{EMP}} (\Phi_1 (U_1(P_1^{\dag_1}, P_2^{\dag_2})))] - E[\Psi_{\text{EMP}} (\Phi_1 (U_1(P^{\dag_1}_1, P^{\dag_1}_2)))].
\]

The incidental gain-loss differential for player 2 is:
\[
\Delta^{M'_2}_2 = E[\Psi_{\text{EMP}} (\Phi_2 (U_2(P_1^{\dag_1}, P_2^{\dag_2})))] - E[\Psi_{\text{EMP}} (\Phi_2(U_2(P^{\dag_2}_1, P^{\dag_2}_2)))].
\]

\subsubsection{Extension: $n$-Player Non-Repeated Coarse-Grained Game with Mixed Strategies and the Entropy-Maximizing Preprocessing} 
In games with three or more players, deriving mixed strategies from the best response requires systematically evaluating each player's strategic choices and ensuring equilibrium conditions. The fundamental approach remains the same as in the two-player case, and the equations presented below are extensions or derivations of those used in the two-player setting. 

First, we introduce the formula for calculating the expected payoff that a given strategy profile yields for player $k$ in an \( n \)-player game.
\begin{align*}
& E[U_k(P_1, P_2, \dots, P_k, \dots, P_n)] = \\
& \quad\quad \sum_{j_1 = 1}^{|S_1|} \sum_{j_2 = 1}^{|S_2|} \dots \sum_{j_k = 1}^{|S_k|} \dots \sum_{j_n = 1}^{|S_n|} \left( \prod_{k=1}^{n} p_{j_k} \right) u_k(s_{j_1}, s_{j_2}, \dots, s_{j_k}, \dots, s_{j_n}).
\end{align*}

\paragraph{Meaning of Each Component}
\begin{itemize}
    \item \( E[U_k (p_1, p_2, \dots, p_n)] \): Expected payoff of player \( k \).
    \item \( S_k \): Strategy set of player \( k \).
    \item \( s_{j_k} \): A pure strategy chosen by player \( k \).
    \item \( p_{j_k} \): Probability that player \( k \) chooses strategy \( s_{j_k} \) (part of their mixed strategy).
    \item \( u_k (s_{j_1}, s_{j_2}, \dots, s_{j_k}, \dots, s_{j_n}) \): Payoff function of player \( k \) when the strategy profile \( (s_{j_1}, s_{j_2}, \dots, s_{j_k}, \dots, s_{j_n}) \) is played.
\end{itemize}

\paragraph{Interpretation of the Formula}
\begin{enumerate}
\item The expression \( \sum_{j_1 = 1}^{|S_1|} \sum_{j_2 = 1}^{|S_2|} \dots \sum_{j_k = 1}^{|S_k|} \dots \sum_{j_n = 1}^{|S_n|} \) iterates over all possible strategy combinations chosen by the players.
\item The term \( \prod_{k=1}^{n} p_{j_k} \) calculates the probability that all players select a specific strategy profile \( (s_{j_1}, s_{j_2}, \dots, s_{j_k}, \dots, s_{j_n}) \).    
\item The product of the probability \( \prod_{i=1}^{n} p_{j_k} \) and the payoff \( u_k(s_1, s_2, \dots, s_{j_k}, \dots, s_n) \) represents the contribution of each strategy profile to player \( k \)'s expected utility.  
\item By summing over all possible strategy profiles, we obtain the expected payoff for player \( k \), considering the randomness introduced by all players' mixed strategies.
\end{enumerate}

Each player's expected payoff is computed by considering the probability distributions over opponents' strategies as in the case of two-player game. For player \( k \), the condition for indifference is:
\begin{align*}
& E[U_k(P_1,P_2,\dots, s_{j_k}, \dots, P_n)] = E[U_k(P_1,P_2,\dots,s_{j'_k},\dots,P_n)], \quad \forall s_{j_k},s_{j'_k} \in S_k.
\end{align*}
This can be expanded as follows:
\begin{align*}
E[U_k(s_{j_k}, P_{-k})] &= E[U_k(s_{j'_k},P_{-k})] \\
\sum_{j_{-k} \in S_k} \left( \prod_{i \neq k} p_{j_i} \right) u_k(s_{j_k}, s_{j_{-k}}) &= \sum_{j_{-k} \in S_k} \left( \prod_{i \neq k} p_{j_i} \right) u_k(s_{j'_k}, s_{j_{-k}}).
\end{align*}

The system of equations is solved to determine the best probability distribution \( P^*_1, P^*_2, \dots, P^*_n \) for each player. This yields the mixed strategy Nash equilibrium, where no player has an incentive to deviate. 

Applying the notation defined for the two-player case, the expected payoff of player \( k \) at Nash equilibrium in the CGG in $M'_k$ is given by the following equation:
\begin{align*}
& E[\Psi_{\text{EMP}} (\Phi_k (U_k (P^{\dag_k}_1, P^{\dag_k}_2, \dots, P^{\dag_k}_k,\dots, P^{\dag_k}_n)))]. 
\end{align*}

The subjective incidental gain-loss differential for each player \( k \) is:
\[
\Delta^{M'_k}_k = E[\Psi_{\text{EMP}} ( \Phi_k (U_k (P_1^{\dag_1}, P_2^{\dag_2}, \dots, P_n^{\dag_n})))] - E[\Psi_{\text{EMP}} (\Phi_k (U_k (P_1^{\dag_k}, P_2^{\dag_k}, \dots, P_n^{\dag_k})))].
\]

\subsection{Unrecognized Gain-Loss Differential}\label{sec5.2}

Objective incidental payoff is defined as the difference between the actual payoff a player receives in the game and the expected payoff they would have received in the base Nash equilibrium. Since we have already established the method for calculating player \( k \)'s subjective payoff in the previous section, we will introduce the formula for objective incidental payoff briefly.

\paragraph{Non-Repeated, Pure Strategy Case}
For an \( n \)-player with a pure strategy game, the objective incidental payoff is given by:
\[
\Delta^{\text{base}}_k = E[U_k(s_1^{\dag_1}, s_2^{\dag_2}, \dots, s_n^{\dag_n})] - E[U_k(s_1^*, s_2^*, \dots, s_n^*)].
\]

\paragraph{Non-Repeated, Mixed Strategy Case}
For an \( n \)-player mixed strategy game, the objective incidental payoff is given by:
\[
\Delta^{\text{base}}_k = E[U_k (P_1^{\dag_1}, P_2^{\dag_2}, \dots, P_n^{\dag_n})] - E[U_k (P_1^*, P_2^*, \dots, P_n^*)].
\]

\section{Nash Equilibrium in Infinite Repeated Game}\label{sec6}
\subsection{Does Folk Theorem Hold in Coarse-Grained Games?}\label{sec6.1}
The considerations so far naturally raise an intriguing question: what happens in the case of repeated games?

Here, let us examine whether the Folk Theorem holds. The Folk Theorem states that in infinitely repeated games, if players sufficiently value future payoffs, a wide range of payoffs can be sustained as Nash equilibria.  

In a standard noncooperative game played once, players typically choose their best response based on immediate payoffs, leading to equilibrium outcomes such as the Nash equilibrium. However, when the game is repeated indefinitely, players can condition their strategies on past actions, allowing for the possibility of long-term incentives.  

Mathematically, in an infinitely repeated game with a discount factor \( \delta \) (where \( 0 < \delta < 1 \)), each player's total discounted utility is given by  
\[
V_k = \sum_{t=0}^{\infty} \delta^t u_k(s_t),
\]
where \( s_t \) is a strategy profile at time \( t \), meaning \( u_k(s_t) \) represents the stage-game payoff of player \( k \) at time \( t \). If \( \delta \) is sufficiently close to 1, meaning players place high importance on future payoffs, they may be able to sustain cooperation under a credible punishment mechanism, as suggested by the Folk Theorem.

The Folk Theorem states that any feasible and individually rational payoff vector can be sustained as a Nash equilibrium, provided that players are sufficiently patient. That is, for any feasible payoff profile \( (V_1, V_2, ..., V_m) (m \in \mathbb{N})\) satisfying  
\[
V_k \geq \bar{U}_k \quad \forall k,
\]  
where \( \bar{U}_k \) is the minmax payoff (the lowest payoff a player can guarantee themselves regardless of opponents' actions), there exists a strategy profile that supports these payoffs in equilibrium.  

A classic example is the infinitely repeated prisoner's dilemma. In the one-shot version, the only Nash equilibrium is mutual defection. However, when the game is repeated, players can adopt strategies such as Tit-for-Tat, where they cooperate initially and mirror their opponent's previous move. Cooperation becomes a Nash equilibrium if \( \delta \) is sufficiently large, as deviating leads to long-term punishment.  

\begin{remark}
A feasible payoff is the set of payoff vectors that can be achieved by some combination of strategies played by all players in the game. Formally, if \( S_1 \times S_2 \times \dots \times S_m \) is the joint strategy space and \( u_k(s) \) is the payoff function for player \( k \), then the feasible payoff set is given by:
\[
F = \left\{ (u_1(s), u_2(s), \dots, u_n(s)) \mid s \in S_1 \times S_2 \times \dots \times S_m \right\}
\]
This represents all possible payoff outcomes that can be realized by some of the players' strategic choices.
\end{remark}

Thus, the Folk Theorem demonstrates how repeated interactions enable cooperative behavior that would not emerge in one-shot games. The key question here is whether this result still holds under coarse-grained perception, and if it holds, how it differs from the standard game.

\subsection{Discount Factor Misalignment in Coarse-Grained Games}\label{sec6.2}
\subsubsection{Fulfillment of the Folk Theorem in Coarse-Grained Matrix}\label{sec6.2.1}
Here, it is necessary to distinguish between the subjective coarse-grained matrix $M'_k$ and the objective base matrix $M^{\text{base}}$. First, in terms of the subjective aspect, the following three lemmata hold (Lemma \ref{subjective_folk_theorem}, Lemma \ref{subjective_discount_rate_discrepancy}, and Lemma \ref{objective_discount_rate_discrepancy}).

\begin{lemma}[Subjective Fulfillment of the Folk Theorem under EMP]\label{subjective_folk_theorem}
Suppose that the EMP is adopted. Then, the Folk Theorem subjectively appears to hold in the coarse-grained matrix \( M'_k \) of any player \( k \). That is, when player \( k \) engages in rational reasoning, it seems that any individually rational payoff could be sustained as a Nash equilibrium from their perspective.
\end{lemma}

\begin{proof}{of Lemma \ref{subjective_folk_theorem}}
We need to show that under the EMP, a player's subjective perception of the coarse-grained matrix \( M'_k \) satisfies the conditions necessary for the Folk Theorem to hold. Specifically, we verify the following conditions:

\begin{enumerate}
    \item Players have perfect information (i.e., they can accurately observe their opponents' choices within their coarse-grained perception).
    \item Payoffs are real numbers (i.e., each strategy profile maps to a well-defined subjective payoff).
    \item The game is infinitely repeated.
    \item Players sufficiently value future rewards (i.e., the discount factor \( \delta \) is sufficiently close to 1).
\end{enumerate}

\textbf{Step 1: Subjective Payoff Representation under EMP.}  
In the coarse-grained matrix \( M'_k \), the payoff perceived by player \( k \) when a strategy profile \( (s_{j_1}, \dots, s_{j_n}) \) is played is given by:

\[
\psi_{\text{EMP}}(\varphi_k(u_k(s_{j_1}, \dots, s_{j_n})))
\]

where \( u_k(s_{j_1}, \dots, s_{j_n}) \) is the true payoff in the base matrix, \( \varphi_k \) is the coarse-graining function mapping real payoffs into a partitioned structure, and \( \psi_{\text{EMP}} \) is the EMP, which assigns a representative numerical value to the coarse-grained partition.

Since the coarse-graining function \( \varphi_k \) partitions payoffs into intervals or sets, and EMP assigns a single representative value per partition, the resulting payoff remains a well-defined real number. This ensures that subjectively, players perceive the game as having well-defined payoffs.

Thus, Condition 2 is satisfied from the subjective perspective.

\textbf{Step 2: Subjective Observability of Opponent Strategies.}  
In a standard perfect monitoring setup, players can observe opponents' strategies exactly. However, in a CGG, player \( k \) only perceives payoffs through \( \varphi_k \), meaning they do not see precisely which strategy was played but instead recognize the outcome at a coarser level.

Since the EMP strategy assigns a deterministic value to each coarse-grained payoff partition, the subjectively perceived game still appears fully observable to the player. In other words, from their perspective, the game behaves as if it were played under perfect monitoring.

Thus, Condition 1 is subjectively satisfied.

\textbf{Step 3: Infinite Repetition and Discounting.}  
The assumption that the game is infinitely repeated is an external assumption that does not depend on coarse-graining. That is, regardless of the level of perception, the game itself remains infinitely repeated.

Similarly, the discount factor \( \delta \) is an exogenous parameter governing how much players value future rewards. The coarse-grained perception does not eliminate \( \delta \), so as long as the player sufficiently values future rewards (\( \delta \to 1 \)), Condition 4 is satisfied.

Thus, Conditions 3 and 4 are trivially satisfied.

\textbf{Step 4: Subjective Perception of Folk Theorem Validity.}  
Since all four conditions of the Folk Theorem are satisfied from the player's subjective perspective, the logical consequence is that the player perceives the Folk Theorem to hold.

That is, player \( k \) believes that any individually rational payoff (greater than or equal to the minmax value in their subjective perception) can be sustained as a Nash equilibrium. 

However, this illusion occurs only because the player perceives a distorted version of the payoff structure. In reality, the base game matrix does not necessarily support all the equilibria the player believes to exist.

Thus, from the coarse-grained perspective of player \( k \), the Folk Theorem appears to hold.
\end{proof}

\subsubsection{Intersubjective Discrepancy in Perceived Sufficient Discount Rates}\label{sec6.2.2}
\begin{lemma}[Intersubjective Discrepancy in Perceived Sufficient Discount Rates]\label{subjective_discount_rate_discrepancy}
Consider an infinitely repeated game where player \( k \) and player \( l \) reason rationally based on their respective coarse-grained matrices \( M'_k \) and \( M'_l \). The sufficient discount rate \( \delta^{M'_k}_l \) perceived by player \( k \) does not necessarily coincide with the sufficient discount rate \( \delta^{M'_l}_l \) perceived by player \( l \).
\end{lemma}

\paragraph{Example: Discount Rate Discrepancy in Coarse-Grained Games}

We present an example demonstrating that, even if player 2 possesses the discount rate that player 1 initially considered sufficient for sustaining cooperation, cooperation may still fail because player 2's perceived sufficient discount rate differs from player 1's expectation.

Consider an infinitely repeated Prisoner's Dilemma with the following base payoff matrix:
\[
\begin{array}{c|cc}
    & \textbf{Remain Silent} & \textbf{Confess} \\ \hline
    \textbf{Remain Silent} & (-1, -1) & (-5, 0) \\  
    \textbf{Confess}       & (0, -5)  & (-3, -3)  
\end{array}
\]

In an infinitely repeated game, cooperation is sustainable if and only if a player $k$'s expected discounted utility from cooperating is at least as high as the expected discounted utility from defecting under an adequate punishment mechanism. 

Mathematically, this condition is expressed as:  
\[
V_k(\text{cooperate}) \geq V_k(\text{defect}),
\]
where \( V_k(\text{cooperate}) \) represents the total discounted utility for player \( k \) if they always cooperate, and \( V_k(\text{defect}) \) represents the total discounted utility if they defect at least once, triggering the specified punishment mechanism. This condition ensures that cooperation is the best response given that other players are also cooperating, thus satisfying the requirements for Nash equilibrium.

We now derive the correct formula for the critical discount factor \( \delta_k \). If the player always cooperates, they receive the cooperation payoff \( T_C \) every period ($T_C = u_k(s_{c_k})$ where $s_{c_k}$ is the strategy that is regarded as cooperating in the current game). The sum of an infinite geometric series gives the total expected discounted utility:
\[
V_k (\text{cooperate}) = \sum_{t=0}^{\infty} \delta_k^t T_C = \frac{T_C}{1 - \delta_k}.
\]

Consider the trigger strategy, where any deviation leads to permanent punishment. If the player $k$ defects once, they receive the temptation payoff \( T_D \) in the first period ($T_D = u_k(s_{d_k})$ where $s_{d_k}$ is the strategy that is regarded as defecting in the current game). However, the opponent responds by permanently defecting, leading the player $k$ to receive the punishment payoff \( T_B \) in all subsequent periods.

Thus, the total discounted utility from deviating is:
\[
V_k(\text{defect}) = T_D + \sum_{t=1}^{\infty} \delta_k^t T_B.
\]

Rewriting the infinite sum,
\[
V_k (\text{defect}) = T_D + \delta_k \frac{T_B}{1 - \delta_k}.
\]

For the trigger strategy to be an equilibrium, the player must prefer cooperation:
\[
V_k(\text{cooperate}) \geq V_k(\text{defect}).
\]

Substituting the expressions for \( V_k(\text{cooperate}) \) and \( V_k(\text{defect}) \):
\[
\frac{T_C}{1 - \delta_k} \geq T_D + \delta_k \frac{T_B}{1 - \delta_k}.
\]

Multiplying both sides by \( (1 - \delta_k) \):
\[
T_C \geq (1-\delta_k)T_D + \delta_k T_B.
\]

Solving for \( \delta_k \), we obtain:
\[
\delta_k \geq \frac{T_D - T_C}{T_D - T_B}.
\]

This is the correct formula for computing the critical discount factor in a trigger strategy equilibrium. In the above case:
\[
\delta_k \geq \frac{T_C - T_D}{T_C - T_B} = \frac{0 - (-1)}{0 - (-3)} = \frac{1}{3},
\]

where:
\begin{itemize}
\item \( T_C = -1 \) is the payoff for mutual cooperation,
\item \( T_D = 0 \) is the temptation payoff (defect while the opponent cooperates),
\item \( T_B = -3 \) is the payoff after the defect.
\end{itemize}

Thus, from player 1's perspective, if player 2's discount factor satisfies \( \delta^{\text{base}}_2 \geq \frac{1}{3} \), cooperation should be sustainable.

Since the payoff matrix is symmetric, the discount factor of player 1 as perceived by player 2 is also \( \delta^{\text{base}}_1 \geq \frac{1}{3} \).

\paragraph{Coarse-Grained Perception and Discount Rate Computation}
Now, suppose the game is played in a coarse-grained setting, where players perceive payoffs in broader categories. The coarse-grained payoff matrices for players 1 and 2, computed using their coarse-grained partitions the EMP, are Table \ref{tb1} and Table \ref{tb2}.

\begin{table}[h]
    \centering
    \caption{Coarse-Grained Payoff Matrix for player 1 ($ M'_1 $)}
    \label{tb1}
    \begin{tabular}{c|cc}
        & \textbf{Remain Silent} & \textbf{Confess} \\ \hline
        \textbf{Remain Silent} & $([-1,0),[-1,0)) \to (-0.5, -0.5)$ & $([-6,-4)),\{0\}) \to (-5, 0)$ \\  
        \textbf{Confess}       & $(\{0\},[-6,-4)) \to (0, -5)$ & $([-4,-1),[-4,-1)) \to (-2.5, -2.5)$  
    \end{tabular}
\end{table}

\begin{table}[h]
    \centering
    \caption{Coarse-Grained Payoff Matrix for player 2 ($ M'_2 $)}
    \label{tb2}
    \begin{tabular}{c|cc}
        & \textbf{Remain Silent} & \textbf{Confess} \\ \hline
        \textbf{Remain Silent} & $([-2,0), [-2,0)) \to (-1, -1)$ & $([-6,-4), \{0\}) \to (-5, 0)$ \\  
        \textbf{Confess}       & $(\{0\}, [-6,-4)) \to (0, -5)$ & $([-4,-2), [-4,-2)) \to (-3, -3)$  
    \end{tabular}
\end{table}

From player 2's perspective, the game still looks like the standard Prisoner's Dilemma, meaning that if both players have a discount factor \( \delta^{M'_2}_1 \geq \frac{1}{3} \) and \( \delta^{M'_2}_2 \geq \frac{1}{3} \), the trigger strategy should sustain cooperation.

However, since player 1's coarse-grained matrix differs from the base matrix, this reasoning does not apply, and the discount factor must be recalculated.

\paragraph{Player 1's Subjective Discount Rate Computation}
For player 1 to find cooperation preferable, the expected discounted value of cooperation under the trigger strategy must exceed the temptation to defect. That is:
\[
V^{M'_1}_2(\text{cooperate}) \geq V^{M'_1}_2(\text{defect})
\]

Solving for the sufficient discount factor \( \delta^{M'_2}_2 \), we get:
\[
\delta^{M'_1}_2 \geq \frac{T_D' - T_C'}{T_D' - T_B'} = \frac{0 - (-0.5)}{0 - (-2.5)} = \frac{1}{5}.
\]

This calculation shows that from player 2's perspective, cooperation is achieved through a trigger strategy when both players have a discount factor of at least \( \frac{1}{3} \), whereas from player 1's perspective, cooperation is achieved when both players have a discount factor of at least \( \frac{1}{5} \). Consequently, if both players have a discount factor of \( \frac{1}{4} \), player 1 expects cooperation through the trigger strategy, while player 2 does not. This discrepancy is an instance of discount factor misalignment.

\paragraph{Theoretical Explanation}
\begin{proof}{of Lemma \ref{subjective_discount_rate_discrepancy}}
We prove that in a CGG, the sufficient discount factor for cooperation may differ depending on the observer's perspective.

In the standard Folk Theorem, cooperation is sustainable if each player's equilibrium payoff $ V_k $ satisfies:
\[
V_k \geq \bar{U}_k, \quad \forall k,
\]
where $ \bar{U}_k $ is the minmax payoff, ensuring that every player receives at least what they could guarantee on their own. 

However, in a CGG, the perceived utility of player $ l $ from the perspective of player $ k $ is given by:
\[
V^{M'_k}_l = \sum_{t=0}^{\infty} (\delta^{M'_k}_l)^t \psi_k (\varphi_k(u_l(s_t))), 
\]
where, as defined in this paper, \( \psi_k = \psi_{\text{EMP}} \). 

For cooperation to be sustained, player $ l $ must believe that deviation leads to lower expected utility from the perspective of $ k $:
\begin{align*}
V^{M'_k}_l (\text{cooperate}) &\geq V^{M'_k}_l (\text{defect}),
\end{align*}
which expands to:
\begin{align*} 
\sum_{t=0}^{\infty} (\delta^{M'_k}_l)^t \psi_{\text{EMP}}(\varphi_k(T_C)) 
&\geq \sum_{t=0}^{\infty} (\delta^{M'_k}_l)^t \psi_{\text{EMP}}(\varphi_k(u_l(s'_t))).
\end{align*}

On the other hand, player \( l \)'s sufficient discount factor from their own perspective is computed as:
\begin{align*}
V^{M'_l}_l (\text{cooperate}) &\geq V^{M'_l}_l (\text{defect}),
\end{align*}
which expands to:
\begin{align*} 
\sum_{t=0}^{\infty} (\delta^{M'_l}_l)^t \psi_{\text{EMP}}(\varphi_l(T_C)) 
&\geq \sum_{t=0}^{\infty} (\delta^{M'_l}_l)^t \psi_{\text{EMP}}(\varphi_l(u_l(s'_t))).
\end{align*}

Now, assuming that the discount factor is not affected by coarse-graining, we obtain:
\begin{align*} 
\sum_{t=0}^{\infty} (\delta^{M'_k}_l)^t \psi_{\text{EMP}}(\varphi_k(T_C)) 
&\geq \sum_{t=0}^{\infty} (\delta^{M'_k}_l)^t \psi_{\text{EMP}}(\varphi_k(u_l(s'_t))) \\
&\iff \sum_{t=0}^{\infty} (\delta^{M'_k}_l)^t \psi_{\text{EMP}}(\varphi_l(T_C)) 
\geq \sum_{t=0}^{\infty} (\delta^{M'_k}_l)^t \psi_{\text{EMP}}(\varphi_l(u_l(s'_t))).
\end{align*}

However, this equivalence does not always hold. Given the definitions of \(\psi_{\text{EMP}}\), \(\varphi_k\), and \(\varphi_l\), there is no guarantee that the value of \( u_l(s_t) \) is preserved under coarse-graining. In particular, \(\varphi_k\) and \(\varphi_l\) may map \( u_l(s_t) \) to different coarse-grained values, leading to discrepancies in the perceived utilities. 

Hence, the lemma is proven.
\end{proof}

\subsubsection{Comparison with the Objective Perspective in $M^{\text{base}}$}\label{sec6.2.3}

\begin{lemma}[Discount Rate Discrepancy in Coarse-Grained Games]\label{objective_discount_rate_discrepancy}
The sufficiently high discount rate \( \delta^{M'_k}_l \) computed based on the coarse-grained matrix $M'_k$ for player $l$ does not necessarily coincide with the sufficiently high discount rate \( \delta^{\text{base}}_l \) computed based on the base matrix $M^{\text{base}}$ for player $l$.
\end{lemma}

Since the validity of this lemma is evident from Section \ref{sec6.2.2}, the proof is omitted.

\subsection{Failure of Cooperation}\label{sec6.3}
Thus, we have established the following: (1) the Folk Theorem can still be subjectively computed in a coarse-grained matrix (Lemma \ref{subjective_folk_theorem}), (2) the sufficient discount rates subjectively computed by different players may also differ (Lemma \ref{subjective_discount_rate_discrepancy}), and (3) the sufficient discount rate computed in the coarse-grained matrix may differ from that computed in the base matrix (Lemma \ref{objective_discount_rate_discrepancy}). Based on these findings, the following theorem holds.

\begin{theorem}[Failure of Cooperation]\label{failure}
In an infinitely repeated game with mixed strategies and EMP, the following two phenomena can occur:
\begin{enumerate}
\item When player \( k \)'s resolution is sufficiently coarse, even if the objective condition for cooperation to be beneficial is satisfied, i.e., the discount factor \( \delta^{\text{base}}_k \) meets the necessary threshold, player \( k \) does not subjectively perceive cooperation as a rational choice.
\item When player \( k \)'s resolution is sufficiently coarse, even if player \( l \)'s perspective suggests that cooperation is objectively beneficial and the discount factor \( \delta^{M'_l}_k \) meets the necessary threshold, player \( k \) does not subjectively evaluate cooperation as the optimal strategy.
\end{enumerate}
\end{theorem}  

A trivial sufficient condition for these cases is:  
\[
\exists G \in \mathfrak{G}_k, \quad \forall j_k, \quad \psi_{\text{EMS}} (\varphi_k (u_k ( s_{j_1}, s_{j_2}, \dots, s_{j_k}, \dots, s_{j_n}))) \in G.
\]
This implies that all payoffs perceived by player \( k \) are mapped to the same coarse-grained set \( G \), making it impossible for the player to distinguish between different strategic payoffs, thus preventing them from recognizing cooperation as beneficial.  

A non-trivial condition for these phenomena is:  

\paragraph{For the First Case:}  
If \( a \) is the minimum discount factor required for player \( k \) to cooperate when calculated using the base matrix, and \( b \) is the minimum discount factor required for player \( k \) to cooperate when calculated using the matrix \( M'_k \), then \( \delta^{M'_k}_k \) satisfies the following condition:  
\[
a \leq \delta^{M'_k}_k < b.
\]  
This indicates that player \( k \)'s perceived discount factor \( \delta^{M'_k}_k \) under the coarse-grained matrix \( M'_k \) deviates from the objectively sufficient threshold in \( M^{\text{base}} \). This discrepancy leads to a misalignment in perceived incentives for cooperation, preventing player \( k \) from recognizing cooperation as the optimal choice.

\paragraph{For the Second Case:}  
If \( a \) is the minimum discount factor required for player \( k \) to cooperate when evaluated under player \( l \)'s coarse-grained matrix \( M'_l \), and \( b \) is the minimum discount factor required for player \( k \) to cooperate when calculated using the matrix \( M'_k \), then \( \delta^{M'_k}_k \) satisfies the following condition:  
\[
a \leq \delta^{M'_k}_k < b.
\]  
This suggests that even if player \( l \) perceives cooperation as beneficial given the threshold \( \delta^{M'_l}_k \), player \( k \) evaluates their own discount factor differently from this threshold, leading to a discrepancy in perceived cooperation incentives.  
As a result, even when cooperation would be optimal under player \( l \)'s perspective, player \( k \) might fail to recognize this due to differences in their perceived intertemporal incentives.

These conditions highlight how coarse-graining affects the perception of intertemporal incentives, creating discount factor misalignment and obstructing cooperation, even when it would be optimal from an objective standpoint.

\begin{proof}{of Theorem \ref{failure}}
The proof of the trivial condition follows from the fact that the payoff player \( k \) receives from the game is fixed at a constant \( c \), regardless of the strategy profile. Consequently, every possible strategy combination is a Nash equilibrium due to payoff homogeneity. Therefore, this situation presents an equilibrium selection problem, meaning that whether player \( k \) chooses to cooperate depends entirely on the focal point.

The proof of the first case of the non-trivial condition follows from the fact that the range of subjectively sufficient discount factors for player \( k \) does not align with the range of objectively sufficient discount factors based on the base matrix, as demonstrated in Lemma \ref{objective_discount_rate_discrepancy}.  

If player \( k \)'s actual discount factor is at least the threshold required in \( M^{\text{base}} \) but falls below the threshold required in \( M'_k \), then objectively, \( k \) should cooperate; however, subjectively, \( k \) does not perceive cooperation as the optimal choice.  

Therefore, a coarse-graining that satisfies the condition \( a \leq \delta^{M'_k}_k < b \) is sufficiently coarse.

For the second case, by applying the same reasoning as in the first case, we establish that the coarse-graining is sufficiently strong to induce the misalignment in cooperation incentives.
\end{proof}

\begin{remark}
This theorem does not state that cooperation in an infinitely repeated game is sustained as expected only when all players have the same resolution. Even if players have different resolutions, their coarse-grained matrices can be identical within the given game.
\end{remark}

\section{Application to Social Science}\label{sec7}
\subsection{Minor Model Change}\label{sec7.1}

In this paper, we aim to demonstrate the applicability of CGGs in social sciences by exploring several modeling approaches in economics. The first analysis focuses on how consumers' coarse-grained perception can influence a firm's decision to implement minor model changes in its products. We formulate the following game:

\begin{itemize}
\item \text{Low-resolution consumer (\textit{C})}: The consumer owns an aging car and is considering purchasing a new vehicle from Firm \( F \). If the two available models appear indistinguishable in terms of quality, the consumer chooses the cheaper option.
\item \text{High-resolution firm (\textit{F})}: The firm offers two models, \( m_1 \) and \( m_2 \), where \( m_2 \) is a minor upgrade of \( m_1 \) with a slightly higher price. The quality of \( m_2 \) is marginally better than that of \( m_1 \), and while its price is also slightly higher, it yields a somewhat better profit margin.
\end{itemize}

This situation can be represented by the following base payoff matrix.
\[
\begin{array}{c|cc}
C ~\backslash~ F & \text{Sell } m_1 \text{ (cheaper)} & \text{Sell } m_2 \text{ (upgraded)} \\ \hline
\text{Buy } m_1 & (5, 6) & (0, 0) \\  
\text{Buy } m_2 & (0, 0) & (5.5, 6.5) 
\end{array}
\]
Suppose that this payoff matrix reflects the resolution of the dealer. Although both (Buy \( m_1 \), Sell \( m_1 \)) and (Buy \( m_2 \), Sell \( m_2 \)) are pure strategy Nash equilibria, creating an equilibrium selection problem, the dealer would expect that (Buy \( m_2 \), Sell \( m_2 \)) will be realized in practice. This is because it serves as a natural focal point where both players maximize their payoffs.the dealer would expect to be able to sell \( m_2 \). 

However, if the consumer \( C \) perceives payoffs through a coarse-grained partition of the form:  
\[
\mathfrak{G}_c = \{ \dots, [5,6), [6,7), [7,8), \dots \}
\]
then, under the application of EMP, the consumer's transformed payoff matrix \( M'_c \) would become:  
\[
\begin{array}{c|cc}
C ~\backslash_{\mathfrak{G}_c + \psi_{\text{EMP}}}~ F & \text{Sell } m_1 \text{ (cheaper)} & \text{Sell } m_2 \text{ (upgraded)} \\ \hline 
\text{Buy } m_1 & (5.5, 6.5) & (0, 0) \\   
\text{Buy } m_2 & (0, 0) & (5.5, 6.5) 
\end{array}
\]
Unlike the previous case, it is not possible to provide the natural focal point to guide the selection. Whether \( C \) decides to buy \( m_2 \) depends on external factors, such as the consumer's preference for new products or their level of trust in the dealer's recommendation.

While the dealer might successfully sell \( m_2 \) under certain conditions, this outcome does not align with their initial rational inference that C chose \( m_2 \) because they gained more payoff from the improvement in the quality of \( m_2 \).

Suppose that Firm \( F \) develops a new model, \( m_3 \). Considering the inclusion of \( m_3 \), we define the following payoff matrix:
\[
\begin{array}{c|ccc} 
C ~\backslash~ F & \text{Sell } m_1 \text{ (cheaper)} & \text{Sell } m_2 \text{ (upgraded)} & \text{Sell } m_3 \text{ (newest)}\\ \hline 
\text{Buy } m_1 & (5, 6) & (0, 0) & (0,0) \\   
\text{Buy } m_2 & (0, 0) & (5.5, 6.5) & (0,0) \\  
\text{Buy } m_3 & (0,0) & (0,0) & (6,7) 
\end{array}
\]

This matrix represents the strategic interaction between the firm and the consumer when three product models are available. The firm can choose to sell only one of the models, while the consumer selects which model to purchase. The payoffs indicate that if the firm offers a particular model and the consumer chooses to buy it, both receive a corresponding benefit. If the consumer attempts to buy a model that is not offered, both receive a payoff of zero.

Next, we analyze how this payoff matrix transforms when considering the consumer's coarse-grained perception.
\[
\begin{array}{c|ccc} 
C ~\backslash_{(\mathfrak{G}_c + \psi_{\text{EMP}})}~ F & \text{Sell } m_1 \text{ (cheaper)} & \text{Sell } m_2 \text{ (upgraded)} & \text{Sell } m_3 \text{ (newest)}\\ \hline 
\text{Buy } m_1 & (5.5, 6.5) & (0, 0) & (0,0) \\   
\text{Buy } m_2 & (0, 0) & (5.5, 6.5) & (0,0) \\  
\text{Buy } m_3 & (0,0) & (0,0) & (6.5,7.5) 
\end{array}
\]

In this coarse-grained payoff matrix, despite the difference in resolution between the dealer and the consumer in perceiving product distinctions, they ultimately reach the same conclusion. Formally, there exist three pure-strategy Nash equilibria corresponding to the diagonal entries of the payoff matrix. However, the strategy profile \((\text{Buy } m_3, \text{Sell } m_3)\) Pareto-dominates the others, yielding strictly higher payoffs for both players. Hence, it stands out as the unique equilibrium under the assumption of payoff-maximizing rationality and mutual optimality selection. Although Pareto dominance may serve as a basis for focal point selection, the equilibrium in this context is driven by payoff improvements resulting from a minor model change that enhances product quality. As such, it differs fundamentally from focal point coordination in settings where such quality improvements are not perceived by the players.

The key point here is that the consumer \( C \) was unable to distinguish the quality of \( m_2 \) but became capable of doing so for \( m_3 \). This shift was not due to the dealer's persuasion or the company's advertising but rather because the quality improvement from the model change exceeded \( C \)'s coarse-grained perception threshold. In other words, the resolution of \( C \)'s perception did not change due to newly acquired information but because the magnitude of the improvement surpassed the boundary at which products were previously indistinguishable. Consequently, the equilibrium selection problem for \( C \) was resolved not through external influence but through an intrinsic shift in the structure of perception.

Based on the above analysis, the following conclusions can be drawn:

\begin{enumerate}
\item \textbf{Equilibrium Selection Problem in Minor Model Changes:} Minor model changes can give rise to an equilibrium selection problem unless the associated improvement in product quality yields a perceived utility gain that exceeds the consumer's coarse-grained perception threshold. When this threshold is not surpassed, consumers are unable to distinguish between the old and new models based on quality, and their choices are instead guided by focal points unrelated to product merit—such as a preference for novelty, the dealer’s persuasive tactics, or the influence of advertising. Conversely, when the perceived gain exceeds the threshold, the quality improvement itself functions as a natural focal point, leading to equilibrium selection driven by product merit rather than extrinsic cues.
\item \textbf{Breakthrough Threshold for Quality-Based Competition:} If a firm intends to compete based on quality improvement, the extent of the model change must be sufficient to generate a perceived utility increase that surpasses the consumer's coarse-grained perception threshold. This type of model change can be interpreted as a breakthrough point---a threshold beyond which the consumer begins to recognize the improvement as meaningful and incorporates it into their decision-making process.  
\end{enumerate}

These findings suggest that a firm's product differentiation strategy must account not only for objective quality improvements but also for how consumers perceive those improvements within their cognitive resolution. Without exceeding the recognition threshold, even significant enhancements may fail to influence purchasing behavior, leaving the decision to be guided by external factors rather than intrinsic product value.

\subsection{Adverse Selection}\label{sec7.2}
Adverse selection is a situation in which one party in a transaction has better information than the other, leading to a market inefficiency \citep{akerlof1970market}. It occurs when the less-informed party cannot accurately distinguish between high-quality and low-quality goods or individuals, distorting decision-making.

In this analysis, we reinterpret \cite{akerlof1970market}'s concept of ``less information'' through the lens of low resolution in a CGG. In the original discussion of adverse selection, the concept of the lemon market is examined. Adverse selection refers to a situation in which information asymmetry in a market causes lower-quality goods or higher-risk transactions to dominate, ultimately leading to a decline in market quality and efficiency.  

A lemon market arises when consumers are unable to clearly distinguish between high-quality used cars (peaches) and low-quality used cars (lemons). In this situation, since consumers cannot accurately assess the quality of a used car, they take the risk into account and are only willing to pay the expected average market price. As a result, sellers of high-quality used cars exit the market, leaving only low-quality cars in circulation. Consequently, the average quality of vehicles in the market declines, reducing overall transaction efficiency.

In this section, we reconstruct the lemon market within the framework of a CGG. We define the players and their strategies as follows:

\begin{itemize}
\item Low-resolution consumer (\textit{C}): A consumer who cannot perceive quality differences within a certain range. For any used cars that fall within this range, the consumer is willing to pay only the average market price, regardless of individual differences in quality.
\item High-resolution dealer (\textit{D}): A dealer who can accurately assess subtle differences in the quality of used cars. Given the price offered by the buyer, the dealer prioritizes selling lower-quality vehicles first.
\end{itemize}

Now, suppose the dealer \( D \) owns both a high-quality used car (peach) and a low-quality used car (lemon) and is negotiating with consumer \( C \), who is looking to purchase a vehicle. The consumer is only willing to pay the average price of peaches and lemons. As a result, if the dealer sells a peach, they incur a loss, whereas selling a lemon yields a profit.  

The coarse-graining in this case is somewhat complex. Fundamentally, what is being coarse-grained is the consumer \( C \)'s perception of the utility of used cars—specifically, the fair valuation of each vehicle. However, the payment itself is not subject to coarse-graining, because it would be implausible for a consumer to be uncertain about whether they paid \$10,000 or \$12,000.  

Thus, the consumer's utility function is given by:  
\[
\psi_{\text{EMP}} (\varphi_C (u_C (s_{j_C}))) - \text{sale price}, \quad (s_{j_C} \text{ is ``Buy } m_1 \text{'' or ``Buy } m_2 \text{''})
\]

where the coarse-graining applies only to their valuation of the car, not to the amount they pay.  

For the dealer \( D \), the utility function is given by:  
\[
\text{sale price} - \psi_{\text{EMP}} (\varphi_D (u_D (s_{j_D}))), \quad (s_{j_D} \text{ is ``Sell } m_1 \text{'' or ``Sell } m_2 \text{''}).
\]

Here, for simplicity, we assume that \( \mathfrak{G}_D \) is the finest possible partition, meaning that the dealer perceives all relevant distinctions in vehicle quality. This assumption allows us to focus on the effects of the consumer's coarse-grained perception.

Now, suppose that the fair valuation of \( m_1 \) is \$20,000, while that of \( m_2 \) is \$10,000. In the original lemon market framework, the amount consumer \( C \) is willing to pay is determined based on the expected value, which is derived from the distribution of car qualities available in the market.  

However, in a CGG, the amount \( C \) is willing to pay depends on their coarse-grained partition. For instance, if \( C \)'s partition is such that they cannot distinguish between cars valued between \$10,000 and \$20,000, then, under EMP, \( C \) evaluates both \( m_1 \) and \( m_2 \) as worth \$15,000. Consequently, \( C \) signals their willingness to pay up to \$15,000 for either car.  

In response, dealer \( D \)'s optimal strategy is to sell \( m_1 \) for \$15,000. However, this holds only if \( D \)'s internal valuation (e.g., acquisition cost) of \( m_1 \) is below \$15,000. If \( D \) values \( m_1 \) at \$15,000 or more, then it would also be rational for \( D \) to decline the sale.

This conclusion aligns with the original insight from the lemon market model, where high-quality products exit the market, leaving only low-quality products in circulation. However, by adopting the framework of a CGG, we can provide a more realistic interpretation in the following ways:  

\begin{enumerate}
\item \textbf{A More Realistic Consumer Evaluation Process:} In Akerlof's original lemon market model, consumer \( C \) makes purchasing decisions based on the market's average price, which assumes an expectation over the distribution of product qualities. However, this assumption becomes unrealistic in extreme cases. For instance, when faced with both an exceptionally low-quality used car and a premium-quality used car, it is implausible for \( C \) to simply rely on the market's average price as a valuation. $C$ would likely recognize the flaws in an extremely low-quality used car and abandon probabilistic reasoning altogether. In a CGG, the consumer's perception follows a predefined partition, which sets realistic limits on their ability to distinguish quality differences.  
\item \textbf{Asymmetry in Recognition Ability:} The problem of adverse selection does not necessarily arise from an asymmetry in the information itself but rather from an asymmetry in the ability to process information. Even if the information is available in the market, consumer \( C \) may still purchase a low-quality used car because they lack the cognitive resolution to differentiate quality effectively. In this framework, the issue is not a lack of information but rather a difference in how information is interpreted, which shifts the problem from one of asymmetric information to one of subjective perception.  
\item \textbf{The Effect of High-Resolution Consumers (\( C' \)) on the Market:} If a high-resolution consumer \( C' \) enters the market—one who can correctly assess the value of a used car at \$20,000---the dealer \( D \) would no longer be able to sell a lemon to them. Importantly, this phenomenon does not occur because \( C' \) has access to more external information, nor because \( D \) has provided additional details. Instead, it happens because \( C' \) possesses a greater ability to interpret the same information accurately. This perspective is difficult to explain within the conventional framework of information asymmetry but can be naturally accounted for using the CGG approach. 
\item \textbf{Market Stability under Resolution Thresholds:} Conversely, if ordinary consumers are able to distinguish between used cars whose fair valuation is below \$10,000 and those above---meaning they can recognize a car that is excessively poor in quality---then dealer \( D \) cannot sell a \$5,000 car over \$10,000. As a result, the endless downward spiral of equilibrium prices predicted in the original lemon market model does not occur, and the market does not collapse. This suggests that, among groups of consumers who share a similar level of perceptual resolution, used cars of appropriately matching quality continue to circulate. For instance, this may help explain why vintage cars tend to be traded primarily among expert buyers.
\end{enumerate}

The above analysis does not claim that previous interpretations of the lemon market were mistaken, nor does it deny the existence of information asymmetry. Rather, the central argument is that subjective constraints---such as coarse perception or limited cognitive resolution---can produce effects similar to those of objective constraints, like information asymmetry.

\section{Conclusion}\label{sec8}

In this paper, we introduced Coarse-Grained Games (CGGs) as a framework for analyzing decision-making under perceptual constraints. While coarse-graining is a standard tool in the natural sciences, it has been largely absent from game-theoretic models in the social sciences. We argue that this omission is problematic because human decision-making is inherently coarse-grained; individuals simplify information when making economic, legal, and political choices.

The key insights derived from the definitions, propositions, and theorems presented in this paper are as follows.
\begin{enumerate}
\item The introduction of coarse-graining into game theory can be formalized using coarse set theory. While this paper does not establish coarse set theory as the only possible foundation for coarse-graining, we demonstrate that games defined within this framework exhibit desirable properties, making them both computationally feasible and analytically tractable.
\item In CGGs, the distinction between subjective and objective perspectives, as well as their comparison, is meaningful. A subjective payoff matrix is obtained based on a player's level of coarse perception—equivalently, their resolution—and can be viewed as a blurred version of the underlying objective payoff matrix.
\item In multi-player settings, where each player may have a different coarse-grained perception, the strategic outcomes can diverge from those derived from a game in which all players act based on the same objective payoff matrix. However, when players adopt entropy-maximizing preprocessing, the Nash equilibria of the CGG do not disappear or become entirely random. Instead, they are transformed in a systematic way. In particular, we highlight the following key results:
\begin{enumerate}
  \item In non-repeated pure-strategy games, a Nash equilibrium in the original game remains a Nash equilibrium in the CGG. However, a strategy profile that was not a Nash equilibrium in the original game may become a Nash equilibrium in the CGG.
  \item In non-repeated mixed-strategy games, if the level of coarse-graining is sufficiently high, the CGG degenerates into a simplified version of the original game.
  \item In infinitely repeated mixed-strategy games, the discount factor that a high-resolution player considers sufficient for sustaining cooperation may be insufficient for a low-resolution player to actually engage in cooperative behavior.
\end{enumerate}
\end{enumerate}

Furthermore, this paper demonstrates that CGGs are applicable to the social sciences by examining two economic phenomena. The first concerns how consumers' coarse-grained perceptions affect their behavior in response to minor model changes in products. A key finding is that when the benefit perceived by consumers from a minor model change is too small, they are unable to distinguish between the old and new versions of the product. This results in an equilibrium selection problem. In such cases, overly subtle model updates fail to generate quality-based competition and instead rely on external focal points---such as consumers' preference for novelty or the persuasiveness of sales strategies---to influence purchasing decisions.

The second theme revisits the classic lemon market problem through the lens of CGGs. The analysis reveals that under certain conditions, the market discourages the sale of high-quality used cars while promoting the sale of low-quality ones. However, this phenomenon is not necessarily due to an objective asymmetry of information, but rather to a subjective asymmetry in consumers' ability to interpret information—that is, a disparity in cognitive resolution.

Together, these two examples illustrate not only that CGGs can be applied to economic modeling, but also that their analytical outcomes often align closely with real-world intuition.

This paper has the following limitations. First, while CGGs were defined using coarse set theory, we have not explored alternative formulations that incorporate more complex structural elements. Second, in mapping sets of real values back to single real numbers, we exclusively employed the entropy-maximizing strategy preprocessing (EMP). Further investigation is needed to assess how different strategy preprocessing methods influence the behavior of CGGs. Finally, our discussion on applications to the social sciences was limited to the formulation of hypotheses rather than empirical validation. We leave these issues for future research.

\section*{Declarations}

\textbf{Funding} \\
This research received no specific grant from any funding agency in the public, commercial, or not-for-profit sectors.

\noindent
\textbf{Conflicts of Interest} \\
The author declares no conflict of interest.

\noindent
\textbf{Ethical Approval} \\
This article does not contain any studies with human participants or animals performed by any of the authors.

\noindent
\textbf{Data Availability} \\
No empirical data was used in this study.

\bibliography{references}

\end{document}